\documentclass[fleqn,usenatbib]{mnras}

\usepackage[T1]{fontenc}

\DeclareRobustCommand{\VAN}[3]{#2}
\let\VANthebibliography\thebibliography
\def\thebibliography{\DeclareRobustCommand{\VAN}[3]{##3}\VANthebibliography}

\usepackage{graphicx}	
\usepackage{amsmath}	
\usepackage{amssymb}
\usepackage{gensymb}

\usepackage{subcaption}
\usepackage{geometry}
\usepackage{float}
\usepackage[font=small,labelfont=bf]{caption}
\usepackage{multicol}
\usepackage{xcolor}

\usepackage{newtxtext,newtxmath}

\newcommand{\GG}[1]{}

\title[A Deep Ocean OOL Scenario]{Sea Ice as an Origin of Life Location for Hycean and Ocean Worlds}

\author[E. F. L. Barrier et al.]{
Edouard F. L. Barrier,$^{1}$\thanks{E-mail: efxlb2@ast.cam.ac.uk}
Nikku Madhusudhan,$^{1}$\thanks{E-mail: nmadhu@ast.cam.ac.uk}
Frances E. Rigby$^{1,2}$
\\
$^{1}$Institute of Astronomy, University of Cambridge, Madingley Road, Cambridge, CB3 0HA, UK \\
$^{2}$ Imperial Astrophysics, Department of Physics, Imperial College London, Blackett Laboratory, Prince Consort Road, London, SW7 2AZ, UK
}

\date{Accepted 2026 September 03. Received 2026 August 07; in original form 2025 November 13}

\pubyear{2026}

\begin{document}
\label{firstpage}
\pagerange{\pageref{firstpage}--\pageref{lastpage}}
\maketitle

\begin{abstract}
Water worlds with potential deep oceans are at the forefront of currently observable exoplanetary habitability. The standard requirements for habitability are the presence of an energy source, a source of nutrients, and a solvent, all of which can be met in exoplanets with deep liquid water oceans. Such planets may be numerous, ranging from hycean worlds and ocean worlds to terrestrial-size planets with enhanced water content. However, abiogenesis requires additional conditions, namely a mechanism to accumulate nutrients to the high concentrations required for prebiotic chemistry, and an energy source to drive the chemistry. A deep ocean, and in some cases the high-pressure ice layers underneath, has traditionally been thought to prevent the required concentrations. Here we outline a location where nutrients could be concentrated in sufficiently high quantities to make abiogenesis possible -- at the interface between the open ocean and the surface ice shelf. Using a 3D General Circulation Model, we explore cases corresponding to several candidate habitable exoplanets which might have dayside sea ice. We discuss how the planet's climate influences the nature and behaviour of the sea ice. We investigate the ability of small-scale ice formation and melting processes to concentrate nutrients to prebiotically useful levels, and also explore the potential for meteoritic impactor fragments to provide a platform for prebiotic chemistry. We find that both of these mechanisms may plausibly provide concentrated chemical feedstock for prebiotic chemistry, which adds an important new perspective when evaluating the prospects of habitability in planets with deep oceans.
\end{abstract}




\section{Introduction}

With the advent of the JWST era, the atmospheric characterisation of exoplanets has extended to ever cooler and smaller planets in the sub-Neptune regime. In particular, the community is beginning to characterise the atmospheres of potentially habitable planets such as K2-18 b, TOI-270 d, and LHS 1140 b \citep{Madhusudhan2023b,Holmberg2024,Benneke2024,Damiano2024,Hu2025,Cherubim2026,Holmberg2026}, all of which have been inferred to be water-rich and, with plausible atmospheric compositions, could have liquid water oceans. Excitingly, there have even been some hints of potential biosignatures \citep{Madhusudhan2023b,Madhusudhan2025a} which are currently being debated \citep{Stevenson2025, Welbanks2026,Pica-Ciamarra2026, Holmberg2026}, increasing the urgency of understanding sub-Neptune habitability in detail.

Water-worlds have emerged as the promising environments for habitability in the sub-Neptune regime. These include Hycean worlds \citep{Madhusudhan2021} and ocean worlds \citep{Leger2004,Kite2018}. Hycean worlds are defined as planets with H$_2$-dominated atmospheres, 10-90\% H$_2$O interior mass fractions, and surface liquid water. We use `ocean worlds' to refer to planets with large water mass fractions and high mean molecular weight (MMW) atmospheres, e.g. CO$_2$- or N$_2$-rich atmospheres.
In both cases their liquid water will take the form of deep global oceans, and these oceans will likely overlay high-pressure ice layers \citep[e.g.][]{Leger2004,Noack2016,Rigby2024a}.

Requirements for a location's habitability have traditionally been threefold. Habitable environments require an energy source (usually radiation from the host star), nutrients (available through a variety of means), and a solvent -- usually liquid water \citep{McKay2014,Cockell2016}. Whether a given exoplanet can be considered `habitable' or in the `habitable zone' is usually evaluated using just the third metric, liquid water availability. This reduces the question of habitability to whether the planet could, for a realistic atmospheric composition, possess a liquid water ocean. For terrestrial exoplanets, the requirement of liquid water is met by assuming a carbon-cycle operating on geological timescales, maintaining a nitrogen-carbon dioxide-water vapour atmosphere of the right size for liquid water \citep[][for a review]{Pierrehumbert2010}. This has led to a wealth of work considering how various factors such as stellar type, tidal-locking, and cloud parametrisations can influence the location of the habitable zone and in particular its inner edge \citep[e.g.][]{Kasting1993, Kopparapu2013,Kopparapu2016,Yang2014,Wolf2017,Turbet2018}. The liquid water requirement is by definition met in the case of Hycean and ocean worlds, provided that their atmosphere has the suitable greenhouse effect to allow a temperate surface, e.g. with the right surface pressure, composition and albedo.

However, the requirement to provide the chemical inventory necessary for life can also reduce the habitability of some planets \citep{Lammer2009, Noack2016, Lingam2018, Journaux2020}. Whilst the standard assumption is that water-rock interactions will provide the required bioessential elements, this will not always be the case. Planets which form with abundant volatiles may have high-pressure ices underlying a deep global ocean \citep[e.g.][]{Seager2007}. This would prevent nutrients from the silicate mantle reaching the surface, with the possible exception of material transported through the ice mantle \citep[see e.g. work such as][]{Choblet2017,Kalousova2018a,Kalousova2018b,Hernandez2022,Lebec2023}. This apparent lack of nutrient transport has led people to categorise water-rich planets to be classified as less habitable or even uninhabitable: they have been sorted into categories such as Class IV habitable \citep{Lammer2009} or H3 habitable \citep{Noack2016}. The silicate crust is not the only source of nutrients, however. \citet{Madhusudhan2023a} found that impactors and initial atmospheric accretion could deliver significant quantities of bioessential elements to a global ocean. This was shown to be enough to raise the concentration of various species in the global ocean to those estimated in the Hadean Earth's oceans, meaning that there could be enough nutrients present to sustain a biosphere, even without transport through the ice mantle.

There is, however, a significant further challenge. Even if such water worlds possess all the required ingredients to maintain and sustain life, they still need to be capable of developing life, i.e. there need to be suitable locations for prebiotic chemistry to occur. There are many possible prebiotic chemistries thought to have occurred on early Earth, with differences in terms of pH, temperatures, energy sources, elemental availabilities or mineral surfaces \citep[see various possibities described in][]{Meadows2020,Sasselov2020,Westall2023}. However, a common requirement in all these possibilities studied is for the various reagents to be concentrated to at least millimolar concentrations or higher, to allow for the efficient synthesis of e.g. nucleobases and amino acids, and for some degree of partitioning to isolate various prebiotic reactions. If the whole planet consists of a single global ocean, especially one with very high pressures and temperatures at the water-ice boundary, then following this logic it would be very difficult to concentrate such nutrients to suitable levels.

This concentration problem seems relevant to a wide range of known exoplanet categories. Several terrestrial exoplanets with characterised masses and radii \citep[for example, the TRAPPIST-1 system,][]{Gillon2017} seem to be less dense than Earth. Whilst this could reflect a core-mantle ratio lower than that of the Earth, it could also point to a volatile enrichment of the planet, which would likely manifest as a global ocean for moderate instellations \citep{Krissansen-Totton2024,Guimond2025}. Although likely in contact with a silicate crust and so possibly hydrothermal vents, the suitability of hydrothermal vents as abiogenesis locations are also a matter of debate, at least on Earth \citep{Seyfried2015,McCollom2016,Klein2019,Moore2021}. In any case, a deep enough ocean will likely suppress volcanic activity and outgassing \citep{Krissansen-Totton2021}, eliminating hydrothermal vents as a possibility. The concentration problem might still be difficult to solve in this case.

The concentration problem is also very relevant to Hycean and ocean worlds. A canonical example of an ocean world is the habitable zone sub-Neptune LHS-1140 b, which recent JWST observations have shown to have a volatile-rich atmosphere \citep{Damiano2024,Cadieux2024b}, whilst it also has a bulk density well below that of Earth \citep{Cadieux2024a}. Recently, \citet{Cherubim2026} also reported  escaping Helium and so detected its atmosphere, although the detection is also consistent with a higher mean molecular weight lower atmosphere. Whilst it may not be an ocean world -- a CO$_2$-N$_2$-supercritical H$_2$O envelope is possible -- it could equally well possess a liquid water ocean overlying high-pressure ices. Many super-Earths and smaller sub-Neptunes, especially ones forming outside the ice line, may well be in a similar state, so this is an important planetary regime to consider. Here there would likely not be any scope for prebiotic chemistry on the ocean floor, and thus, in the conventional understanding, no way of prebiotic chemistry to occur. 

The problem of concentration is similarly important for Hycean planets. For example, JWST observations have been interpreted as showing that the sub-Neptune K2-18b may be a Hycean world \citep{Madhusudhan2023b,Hu2025}. If so, it would have high-pressure ice layers in between a liquid water ocean and the silicate interior \citep{Rigby2024a}, leading to the same difficulties in providing a solid substrate for prebiotic chemistry.

Given these difficulties, providing new possibilities for abiogenesis on planets with deep water oceans would significantly change our understanding of their habitability. In this work, we outline the potential for surface sea ice to provide a location for prebiotic chemistry. The core idea is that many habitable zone exoplanets may possess dayside surface sea ice sheets. The presence of such a solid substrate in a habitable zone planet with a deep ocean is already a significant feature which would allow -- as a general principle -- for feedstock molecules to be concentrated. The fact that this substrate can be located on the dayside of the planet's surface is also important, as it provides a natural energy source of the prebiotic chemistry: the incoming stellar irradiation.

To be able to evaluate the feasibility of this abiogenesis location, we should understand the characteristics of sea ice on Earth. For water with an Earth ocean-like salinity, ice formation begins when a stretch of ocean cools below $\approx 271.29$~K, $1.86$~K below the freezing point of pure water \citep{Thomas2010,Fletcher2009}. As seawater's would-be maximum density is below its freezing point, the entire mixed layer will experience convective overturning and is cooled to the freezing point before freezing starts. Ice crystals form around impurities in the upper water layers in the shape of needles, spicules or platelets, known as frazil ice \citep{Weeks1986}. They are kept suspended in the water until a surface layer of ice slush builds up, reducing wind-driven mixing, gradually congealing into granular ice. 

Sea ice freezes as ice Ih. Only a very few other ions (such as fluorine and ammonium ions) are incorporated into the lattice, and most common species present in salt water (Na$^{+}$, K$^{+}$, Ca$^{2+}$, Mg$^{2+}$, Cl$^{-}$, SO$_4^{2-}$, CO$_3^{2-}$) are rejected into the water beyond the ice-water interface, increasing its salinity. Loose masses of frazil crystals have ice volume fractions of 10-30\%, and as they consolidate and expand there remain large numbers of seawater inclusions with considerably enhanced salinities. As the freezing point of seawater decreases with increasing salinity, these inclusions remain liquid even when substantially colder than 271~K.

How sea ice growth continues depends on the physical characteristics of the environment. In more turbulent conditions -- as commonly found in the Antarctic -- the dynamic ice-growth regime leads to frazil ice accreting into pancake ice, decimetre-sized pans of ice with raised edges, which will eventually consolidate \citep{Wadhams1987}. In more quiescent conditions, a surface layer of granular ice is thickened by congelation ice forming below. Here ice growth initially starts planar at the ice-water interface, with salt rejection from the ice matrix forming a thin saline layer at the ice-water interface with a lower melting point. As heat transport is faster than brine transport, there can exist a constitutionally supercooled layer below this, which favours the growth of whichever microscopic structures reach this layer first \citep{Wettlaufer1998}. This leads to lamellar (parallel layers) or dendritic (`tree-like') growth of ice crystals, with salinity-enhanced brines trapped in the spaces between the ice structures.

Sea ice consistently forms with a high porosity (values well above 0.5 are common in new ice) and is filled with brine inclusions \citep{Thomas2010}. This porosity will decrease as the sea ice continues to evolve, getting older and thicker (and colder, during the winter months). Freezing out of the pore space increases the salinity of the brine, but there is also substantial brine movement downwards through the various channels in the ice, resulting in a decrease in the ice's bulk salinity \citep{Eicken1992}. This in general continues until pores are disconnected from each other: this percolation threshold seems to be about at a porosity of about 0.3, and experimental observations suggest that desalination stops when porosity reaches 0.05-0.07 \citep{Cox1988,Arrigo1993}. To give some indicative values, 1 m$^3$ of sea ice will usually contain $10^{14}-10^{15}$ brine inclusions about $10-100\ \mu$m in diameter \citep{Trinks2005}. Additionally, small crystals of salt concentrate out as their solubility limit is reached \citep{Assur1960}. The brine formation process can also lead to pH \citep{Bronshteyn1991} and electric potential gradients \citep{Trinks2005}.

The mechanics of the melting of sea ice also have an impact on its salinity and hence on the feasibility of concentrating nutrients. If the air temperature and/or solar radiation are high enough, melting of the ice will happen top down \citep[e.g.][]{Untersteiner1968}. Meltwater ponds on the ice surface reduce the albedo of the ice-water system, increasing absorption of stellar flux and favouring further melting. A large amount of meltwater percolates through the ice sheet, flushing out brine and substantially reducing the bulk salinity. Melting on the edge and underside of the ice sheet is also common -- in this case increased water temperatures, usually also due to solar radiation, raise the temperature at the ice-water interface above freezing. This causes the bottom of the ice volume to simply melt into the ocean, and the evolution of other portions of the ice volume is largely dictated by their own temperature evolution.

On Earth, the formation and melting of sea ice follows, predictably, a strong seasonal pattern. The Antarctic ice front advances by about 2500~km over its winter, and first-year ice usually reaches a thickness of about 0.5-1.5~m \citep{Wadhams2014,Haas2008}, with the final depth mostly controlled by the upwards ocean heat fluxes. Dynamics (which here we take to mean the small- and large-scale motions of the atmosphere) also has an important impact on ice thickness and distribution. Rough conditions which lead to the formation, movement, and collision of bodies of ice lead to occasional ice thicknesses far greater than predicted from simple thermodynamic ice growth, with deep keels exceeding 10 m in depth \citep[e.g.][]{Fukamachi2006}. Larger-scale wind and ocean gyres can favour the formation of thicker, older ice by keeping the ice in stable, cold conditions, while divergent winds in a given location can encourage ice drift into lower latitudes \citep{Thomas2010}.

The differences between Arctic and Antarctic sea ice are illuminating, as they highlight the impact that a variety of parameters can have on average sea ice characteristics.
Due partly to a less stratified ocean, ocean heat fluxes in the Antarctic are a order magnitude higher \citep{Krishfield2005,Lytle2000} and can reach 100 Wm$^{-2}$ \citep{Muench2001}, which results in a halving of the average year-ice thickness relative to Arctic averages. This is somewhat mitigated by increased moisture availability leading to higher snow cover, which allows water to freeze onto the top of the ice sheet \citep{Maksym2000}.
Differences in large-scale circulations which also have a large impact. Ocean and so ice motion in the Arctic is constrained by the North American and Eurasian landmasses, leading to convergent ice motion, longer residence times and thicker, more deformed ice. 
Ice drift in Antarctica is instead mostly divergent, moving existing thickening ice northwards \citep{Kottmeier1992}. 
Conversely, the atmospheric and oceanic flows of the Antarctic Circumpolar Current act to shield Antarctic sea ice from warm air from lower latitudes, keeping the Antarctic atmosphere colder and drier. Combined with the increased distance from other continents meaning fewer surface impurities, very little surface melting happens in Antarctica, with under-ice melting dominating \citep{Nicolaus2006}.

\vspace{0.2cm}

This example of Earth can give us some initial insights into sea ice's plausibility as a prebiotic chemistry substrate, but we expect that sea ice might act quite differently on exoplanets, and so establishing its behaviour is a crucial objective. More specifically, we are interested in when, how, and where sea ice may be present on various types of planets, in its thickness distribution and evolution, and in what its large-scale formation and melting processes look like. We are also interested in atmospheric conditions such as the surface temperatures, winds, and their variability, all of which strongly impact sea ice behaviour. 

To study these questions, a powerful tool is a General Circulation Model (GCM), which simulates the planet's atmospheres and surface in 3D. Using a GCM allows us to self-consistently study the climates of these planets -- including any sea ice -- and so to investigate the properties of dayside sea ice, and of the wider climate system that influences it. We carry out simulations using the ExoCAM GCM, which can simulate a wide range of terrestrial exoplanet and temperate sub-Neptune atmospheres \citep{Wolf2022,Barrier2025a}. We study three planets across the terrestrial to sub-Neptune range, chosen to represent a range of possible water-worlds. 

We note that the concept that surface sea ice could provide a substrate for prebiotic chemistry is not new, and has been explored in an Earth context by a variety of studies \citep[e.g.][ we review these in Section \ref{sec:prebiotic-chemistry}]{Vajda1999, Trinks2005, Monnard2008a, Attwater2010,Feller2017}. This research has mostly focused on the in-ice feasibility of different stages of prebiotic chemistry, ranging from initial formation of amino acids from a frozen ammonium cyanide solution \citep{Levy2000, Miyakawa2002b} to the replication of RNA in sea ice \citep{Attwater2010,Attwater2013}. Equally, the relevance of this possible mechanism to exoplanets was put forward by \citet{Ramirez2018}, who focused on the `ice habitable zone' by considering the parameter space in which a CO$_2$ dominated atmosphere would be in equilibrium with surface CO$_2$ clathrates. However, there are important differences between these two works. Firstly, we focus on the prebiotic chemistry potential of a wider range of planets, particularly Hycean planets, and also investigate its application to terrestrial planets with ocean-silicate interfaces. Secondly, instead of using a 2D Energy Balance Model (EBM) to analyse the parameter space in which a carbon cycle is operational, we use a full 3D GCM and focus on the behaviour and presence of sea ice under a wide range of planet types. 

Finally, in this work we also consider in some detail possible accumulation mechanisms ways in which elemental feedstocks necessary for prebiotic chemistry might accumulate. This is an important question, as it is not guaranteed that such mechanisms exist. We consider two initial ways: first, we investigate whether small-scale ice formation and melting processes might allow for some sort of freeze-thaw cycle and nutrient concentration from the initial dilute global ocean. Secondly, we consider the possibility that small asteroids or comets might break apart in the atmosphere and deliver small (pebble-sized) material to rest on top of the surface ice sheets. This would lead to regions of the surface with a very concentrated supply of the required prebiotic elements and molecules. This is somewhat similar to \citet{Walton2024}, although instead of small impactors they consider cosmic dust which is concentrated by the atmospheric circulation, and also vaguely similar to the `warm comet pond' idea as suggested by \citet{Pearce2017}.  Either of these mechanisms -- or some other unexplored mechanism -- could lead to regions that have feedstock molecules, an energy source (the host star's energy), a solvent in the form of liquid water, and a solid substrate to concentrate nutrients on, and so are canonically habitable and plausible locations for prebiotic chemistry. 

In Section \ref{sec:methods} we present an overview of the computational models used. In Section \ref{sec:gcm} we present our GCM results. In Section \ref{sec:seaice} we explore in more detail the plausibility of ice sheets as origin of life locations and the feasibility of particular mechanisms to concentrate nutrients. Finally, in Section \ref{sec:summary} we summarise and discuss our findings.

\section{Methods: the ExoCAM GCM}

\label{sec:methods}

In this section we describe the General Circulation Model we use in this work. More comprehensive overviews of ExoCAM and aspects of our implementation can be found in \citep{Wolf2022,Barrier2025a,Barrier2025b}. Here we focus on pertinent details of the atmosphere, ocean and sea ice models used.

\subsection{Atmosphere Model}

We use a modified version of the ExoCAM General Circulation Model \citep{Wolf2022, Barrier2025b} to study the 3D structure of the climates we find, and in particular the presence of behaviour of sea ice. ExoCAM is a specially developed GCM for use in exoplanet settings, and is an independently curated branch of NCAR's Community Earth Systems Model (CESM), version 1.2.1 \citet{Neale2010}. It has been used for a range of terrestrial and more recently sub-Neptune applications \citep[e.g.][]{Haqq-Misra2018, Sergeev2022,Barrier2025a,Barrier2025b}.

The ExoCAM dynamical core uses a finite-volume scheme \citep{Lin1997} to solve the primitive equations of meteorology \citep{Vallis2017}. Angular momentum conservation is a known issue for the finite-volume core on slow-rotator planets, especially ones without significant orography \citep{Lebonnois2012, Lauritzen2014}. To fix this, we employ the corrections of \citet{Toniazzo2020}, who found that the principal angular momentum sink occurred when solving the shallow water equations. 

We run our simulations with a horizontal resolution of 72 by 45 cells, equivalent to cells of $5\degree$ longitude by $4\degree$ latitude. In our H$_2$-dominated atmosphere cases, we use 61 vertical levels spanning 4 orders of magnitude so that the surface is at 1 bar and the model top at 10 Pa. These vertical layers are mostly equally spaced in log pressure, but are concentrated towards the near-surface layers. This gives us the required vertical resolution to resolve near surface effects, notably the presence of a stable non-convecting layer \citep{Innes2023, Seeley2025, Barrier2025b}. In the non-H$_2$ atmosphere cases, we instead use 51 vertical levels spanning 5 orders of magnitude.

We use the ExoRT \textit{n68equiv} radiative transfer scheme as detailed in \citet{Wolf2022}. It is a two-stream correlated-k distribution scheme with 68 spectral intervals from 0.238-infinity $\mu m$. It includes gas absorption for $H_2O$, $CO_2$, $CH_4$, CIA from $N_2-N_2$, $H_2-H_2$, $CO_2-CO_2$, $CO_2-H_2$, and $CO_2-CH_4$ pairs, Rayleigh scattering from $N_2$, $CO_2$, $H_2O$, and $H_2$; finally, liquid and ice cloud droplets are treated as Mie scattering particles. The ExoRT scheme is thus capable of handling realistic radiative transfer for a large number of possible atmospheres, including the N$_2$, CO$_2$, and H$_2$ atmospheres we will consider here.

The CAM 4 moist physics package \citep{Neale2010} predicts the water vapour, liquid cloud, and ice cloud concentrations using a Sunqvist-style bulk microphysical parametrisation \citep{Sundqvist1988, Zhang2003}. There are three types of cloud diagnosed: marine stratus clouds, convective clouds (which we only allow to exist if water has condensed in the convective scheme), and layered cloud if the grid-box relative humidity (RH) is high enough. Convection is controlled by the \citet{Zhang1995} deep and shallow \citep{Hack1993} moist convection schemes, which have been extensively reworked to handle a range of convection types expected in the sub-Neptune regime  \citep{Barrier2025a}. Notably, this allows us to model moist convective inhibition \citep{Guillot1995,Leconte2017}, which is expected on many Hycean planets \citep{Innes2023,Leconte2024,Seeley2025,Gao2025}.

In some of the simulations, we include a parametrisation of photochemically produced hazes, which are expected to be present in shallow hydrogen-rich atmospheres with liquid water oceans \citep[e.g.][]{Huang2024}. We use a simple parametrisation of increasing the optical depth of H$_2$ Rayleigh scattering, fully described in \citet{Barrier2025b}. In this work we treat the hazes as having Rayleigh-like scattering slopes ($\sigma(\lambda) \propto \lambda^{-4}$), and vary the enhancement factor $E$ of the hazes, with $E=1$ corresponding to a clear atmosphere.

\subsection{Ocean and Sea-Ice Models}

We follow ExoCAM standard practice and do not include the POP dynamic ocean model \citep{Smith2010} in our simulations. In our case, this decision is motivated by the fact that critical parameters we would have to set, such as the salinity and the ocean depth, are unknown and could span a large range of values \citep[for some works investigating the effects of this, see][] {Olson2020,Olson2022,Lai2022,Batra2024}. The spin-up times of dynamical oceans are also prohibitive, commonly taking thousands of Earth years of simulation time, and likely substantially longer than that in our thicker atmosphere cases where the wind and stellar forcings would be weaker. Given that we are focused on the possible presence and behaviour of sea ice across a range of planetary paradigms, we leave the study of ocean circulations to future work. Instead we use a slab ocean model, which treats each ocean grid cell as being a single layer in contact with the ice and/or the atmosphere, and a temperature evolution dictated by its interaction with the above components.

We use a fully parametrised sea ice model, CICE \citep{Hunke2008}. This is a 1D ice model that accounts for the presence of multiple layers of snow, pond water, and ice, and solves for the evolution of ice thickness distribution in both space and time. This simulates the thermodynamic changes in ice thickness due to temperature changes and rainfall, assuming a fixed salinity profile matching Earth values for older multiyear ice \citep{Bitz1999}. The bottom of the ice sheet is kept at the freezing point of seawater (-1.8\degree$ C$ in our model), and the ice top cannot exceed 0$\degree C$. However, as we are not using a dynamic ocean model, the parametrisation of ice dynamics, sea ice drift, and ice rifting is not active. This will have an impact on our results, and will in general somewhat increase the open water fraction \citep{Yang2020}. We discuss the various impacts that our sea ice and ocean parametrisations have throughout Sections \ref{sec:gcm}, \ref{sec:seaice}, and \ref{sec:summary}.

\subsection{Simulations performed}

\begin{table*}
\centering
\begin{tabular}{|c|c|c|c|}

\hline \hline
 & \textbf{Case 1} & \textbf{Case 2} & \textbf{Case 3} \\
Analogous planet? & TRAPPIST-1 e & LHS-1140 b & K2-18 b \\ 
\hline
\textit{Bulk planet parameters} & & & \\
$R/R_{\oplus}$ & 0.92 & 1.73 & 2.61 \\ 
g / $ms^{-2}$ & 7.23 & 18.4 & 12.4 \\ 
$P_{day}/$ Earth days & 6.1 & 24.73 & 32.94 \\ 
$P_{orb}/$ Earth days & 6.1 & 24.73 & 32.94 \\ 
Planet obliquity $/ \degree$ & 0 & 0 & 0 \\
Instellation $/S_{\oplus}$ & 0.66 & 0.42 & 1.01 \\
Instellation $/Wm^{-2}$ & 900 & 572 & 1370 \\ 
\hline
\textit{Stellar parameters} & & & \\
Stellar profile & BT-Settl, TRAPPIST-1 & BT-Settl, LHS 1140 & BT-Settl, K2-18 \\ 
Stellar $T_{eff}$ & 2600 & 3100 & 3500 \\
Stellar [Fe/H] & 0 & 0 & 0 \\
\hline
\textit{Atmospheric parameters} & & & \\
Dry N$_2$ VMR & 0.996 & 0 & 0 \\ 
Dry CO$_2$ VMR & 0.0004 & 1.0 & 0.00045 \\ 
Dry CH$_4$ VMR & 0 & 0 & 0.0871 \\ 
Dry H$_2$ VMR & 0 & 0 & 0.77938 \\ 
Dry He VMR & 0 & 0 & 0.13307 \\ 
H$_2$ Rayleigh Scattering enhancement & 1 & 1 & 400 \\
\hline
\textit{Surface parameters} & & & \\
Surface pressure $/$bar & 1 & 5 & 1 \\ 
Internal temperature $/ K$ & 30 & 30 & 30 \\ 
Ocean albedo & 0.06 & 0.06 & 0.06 \\ 
Slab ocean depth $/m$ & 100 & 100 & 100 \\ 
Ocean salinity $/$ psu & 35 & 0 & 35 \\
\hline
\textit{Simulation Parameters} & & & \\
Model top pressure $/$ Pa & 1 & 5 & 10 \\ 
Atmosphere layers & 51 & 51 & 61 \\ 
Runtime $/$ Earth days & 9500 & 19500 & 55000\\ 
\hline \hline
\end{tabular}
\caption[Simulation details of GCM runs]{Details of the GCM simulations carried out. Listed are the planetary, stellar, atmospheric, and simulation parameters. }\label{tbl:gcm_runs}
\end{table*}

We are interested in the possible characteristics of sea ice for a range of planet types. Here, as opposed to \citet{Ramirez2018}, we do not set out to impose an `ice cap zone' analogous to the conventional habitable zone, where surface CO$_2$ clathrates might result in sea ice with a stable climate. As we are interested in several categories of planets with many different possible parameters to span, such a task would carry an immense computational cost, and establishing such a zone using 3D models is well beyond the scope of the paper. In addition to this, the behaviour of carbon cycles and the possibility of stabilising loops with other greenhouse gases, e.g. methane, is still unclear for the range of planets we look at  \citep[e.g.][]{Kitzmann2015,Hakim2023,Levi2017,Levi2019} and we choose not to make assumptions or parametrisations based on this.

Instead, we run GCM simulations of planets' atmospheres spanning a range of compositions and parameter spaces, all of which would have a global water ocean. The aim of this is twofold. Firstly, we show that many planets which possibly have deep oceans would also have dayside sea ice with plausible atmospheric compositions. Secondly, we investigate the nature of the sea ice (extent, thickness, time variability) and that of the surface atmospheric conditions above it (winds, temperatures, time variability). Specifically, we run 3 cases outlined in Table \ref{tbl:gcm_runs}, of which 2 have secondary atmospheres (N$_2$,CO$_2$, or H$_2$O dominated) and 1 has an H$_2$-dominated primary atmosphere, making it a `Hycean' atmosphere.

We simulate two secondary atmospheres corresponding to plausible atmospheric compositions for TRAPPIST-1 e and LHS-1140 b respectively. Both of these planets have sub-Earth densities, and may have significant volatile fractions. On TRAPPIST-1 e, this fraction is high enough that any ocean would likely be global (no continental land mass) if an Earth-like silicate-iron ratio is assumed \citep{Gillon2017}. On LHS-1140 b, the bulk density constraints mean any ocean would be global and accompanied by a significant layer of high pressure ice \citep{Cadieux2024a}. For TRAPPIST-1 e, we adopt an Earth-like atmosphere: 1 bar of N$_2$ with 400 ppm of CO$_2$, matching the simulation parameters of the THAI Hab 2 case \citep{Fauchez2020,Sergeev2022}. In this and other cases, we also include H$_2$O in addition to the background atmosphere: the concentration of this is tracked locally and is governed by evaporation from the surface, advection, and condensation when appropriate. For LHS 1140 b, we assume a thicker atmosphere with 5 bar of CO$_2$ as in \citet{Cadieux2024a}, noting that molecules with $\mu>9$ would not be expected to be entrained by the He outflow and so that a CO2-dominated atmosphere would be consistent with observations \citep{Cherubim2026}. Both of these planets are likely tidally locked. LHS-1140 b is further from its host star, but any high pressure ice layers are expected to lead to a low tidal quality factor Q \citep{Tobie2019} and so leave it tidally locked.

We also model a hycean case with a primary atmosphere, corresponding to the sub-Neptune K2-18 b. In this case, we suppose that it has a thin (1 bar) atmosphere. We use the median retrieved CO$_2$ and CH$_4$ abundances from \citet{Hu2025}, assume a solar H$_2$-He split for the background gas, and apart from a variable H$_2$O component do not include additional gas species. Photochemical modelling of cool hydrogen atmospheres has shown that H$_2$O condensation will leave a carbon-rich stratosphere, which leads to longer chain hydrocarbons and photochemical hazes \citep{Madhusudhan2023a,Huang2024}. For this reason, we include a moderate Rayleigh scattering enhancement of 400x, where a 1x enhancement represents a clear atmosphere. This enhancement value was informed by \citet{Barrier2025b}, and was chosen so that it would lead to partial dayside sea ice coverage.

We choose internal temperatures of 30~K for the sub-Neptune atmospheres, following considerations from \citet{Valencia2013}, and also for the TRAPPIST-1 e case. This corresponds to a heat flux of $\approx 45 $ mW m$^{-2}$, somewhat less than Earth's \citep[e.g.][]{Davies2010}. In any case, internal heating is negligible given the atmosphere sizes we simulate. Our slab ocean model uses a single layer with a depth of 100 m, corresponding to the depth of the wind-mixed layer on Earth. All of our simulations are initialised with an at rest, isothermal atmosphere and surface at 300~K. We consider that the climate states have converged when the kinetic energy of the atmosphere reaches a steady value, the net emergent longwave flux reaches a steady value such that the OLR is equal to the incident stellar irradiation (with a difference  $<1$ Wm$^{-2}$), and the globally-averaged surface temperature is steady. This takes different amounts of time for the different runs, explaining the variations in final runtime. The results shown here are taken for the last (Earth) year of the simulations, with the instantaneous value of the variables measured every 24 hours.

The ocean and ice parameters are kept constant between runs, with the exception of the ocean salinity. This influences the ocean's freezing point: at an Earth-like salinity of roughly 35 psu, the freezing point of ice is reduced by $1.86$ K. However, there is no reason that ocean worlds would have Earth-like salinity, and it can be instructive to consider different ocean salinities. For Case 2 (LHS-1140 b analog), we then assume a minimally saline ocean with no freezing point depression.

\section{GCM Results}
\label{sec:gcm}

In this section we present the results of our GCM simulations investigating the presence of sea ice in a range of planet types, and in particular we describe the behaviour of sea ice and of the main climate features affecting it. We show that a wide range of atmosphere types and sizes, corresponding to several keystone exoplanets, can plausibly have dayside sea ice. We present results of three simulations: Case 1 corresponds to TRAPPIST-1 e, Case 2 to LHS 1140 b, and Case 3 to K2-18 b.

To give some insight into the global climate state, we begin by showing the atmospheric thermal structure. We then focus on the conditions directly relevant to the sea ice: the surface temperatures and their variability, the sea ice fraction, the distributions of sea ice thicknesses, freezing and melting rates, and then the time variability of surface conditions in specific locations. We finish by examining the atmospheric winds, the cloud distribution, and the stellar radiation reaching the surface, all of which will affect the plausibility of sea ice prebiotic chemistry.

\subsection{Atmospheric Thermal Structure}

\begin{figure}
    \centering
    \includegraphics[width=0.48\textwidth]{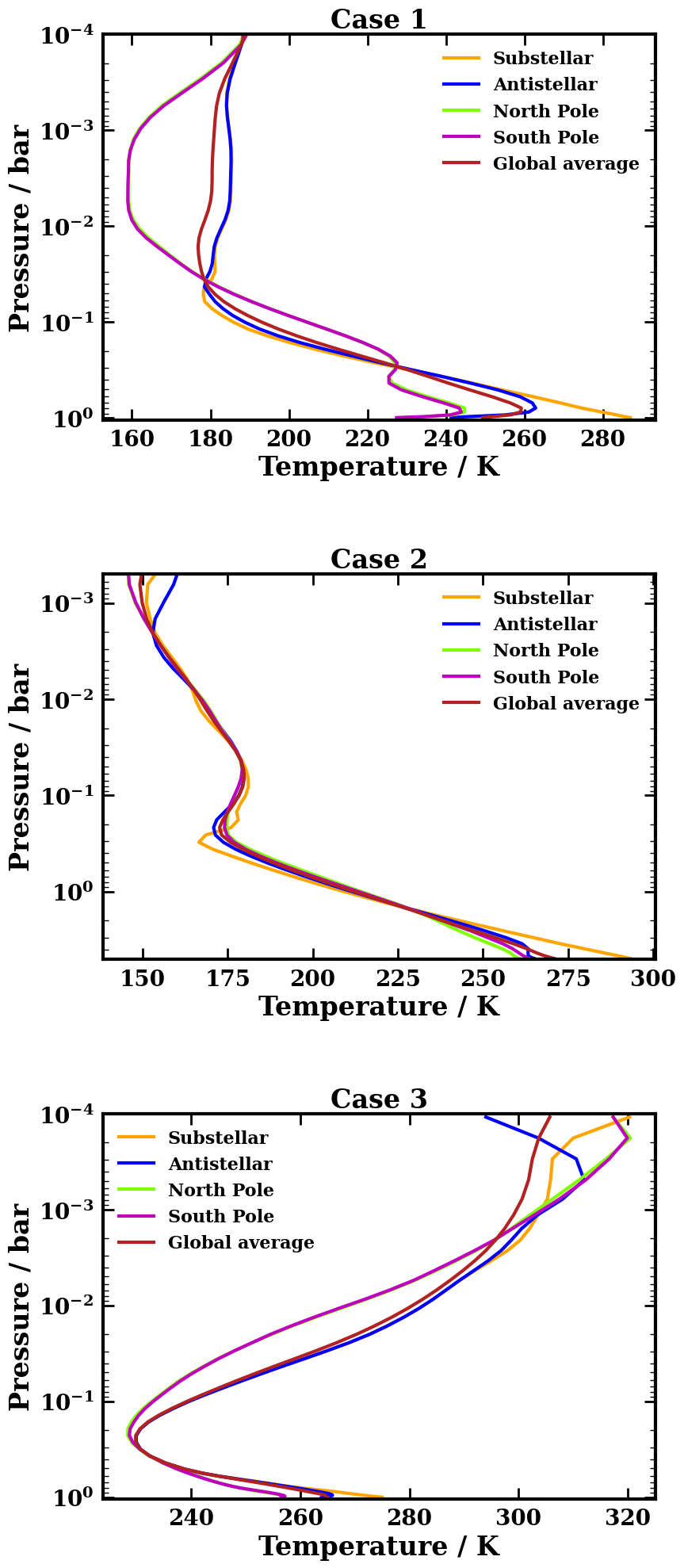}
    \caption{Pressure-temperature profiles of our GCM simulations. By design, all of our surface global temperature averages are near freezing. Depending on the atmospheric composition assumed, there are substantial vertical variations in temperature: see the large temperature inversion, caused mostly by CH$_4$, in Case 3. The magnitude of surface temperature differences varies case by case. For smaller, less optically thick atmospheres e.g. Case 1, there are differences of $>50$~K. As the optical thickness of the atmosphere increases, the surface temperature differences reduce.}
    \label{fig:PTs}
\end{figure}

One of the most important features of planetary climates is the atmospheric thermal structure. We show the atmospheric pressure-temperature (PT) profiles in Figure \ref{fig:PTs}. We include the PT profiles at the substellar and antistellar point, the north and south poles, and a global average. By construction, all of these profiles have moderate surface temperatures, with the global temperature averages slightly below freezing. 

We can note some effects of the different atmospheric compositions. The  lower the optical depth of the atmosphere, the greater the surface temperature differences: Case 1 has the highest surface temperature differences, followed by Case 2, followed by Case 3.  This is because less stellar flux reaches the surface and near-surface regions, so there is less energy to drive horizontal temperature differences. Even though Case 2 has a 5 bar atmosphere, the optical depth of the CO$_2$ atmosphere is still lower than the 1 bar, H$_2$-dominated atmosphere of Case 3. The temperature differences are also affected by the slow rotation rates of the tidally locked cases. Particularly for Cases 2 and 3, the slow rotation rates push them into what is referred to as the weak temperature gradient (WTG) regime \citep[e.g.][]{Pierrehumbert2016} where atmospheric circulation efficiently transports heat and leads to small horizontal temperature differences in the free (above planetary boundary layer) atmosphere. 

There is a significant temperature inversion in Case 3. This is caused by various shortwave absorbers, notably the high CH$_4$ abundance and the parametrised increased Rayleigh scattering. The increased CH$_4$ means that the atmospheric composition observed for K2-18~b by \citet{Hu2025} leads to a noticeably larger anti-greenhouse effect than the atmospheric composition measured in \citet{Madhusudhan2023b}. The influence of the CH$_4$ in causing the temperature inversion can be seen by comparing Case 3, with a 400x Rayleigh scattering enhancement factor and an average surface temperature of 265~K, to the 1b-500x case from \citet{Barrier2025b} with an average surface temperature of 320~K. We expect this to have important implications for the critical albedo required to stop a runaway greenhouse on a Hycean K2-18~b. The increased CH$_4$ abundance means that a lower haze enhancement and hence a lower albedo \textbf{($\gtrsim 0.2$)} is required for a stable atmosphere compared to the constraints in \citet{Barrier2025b}. This, in turn, will increase the maximum possible atmospheric thickness consistent with observations.

\subsection{Sea Ice Presence and Surface Temperatures}

\begin{figure*}
    \centering
    \includegraphics[width=0.95\textwidth]{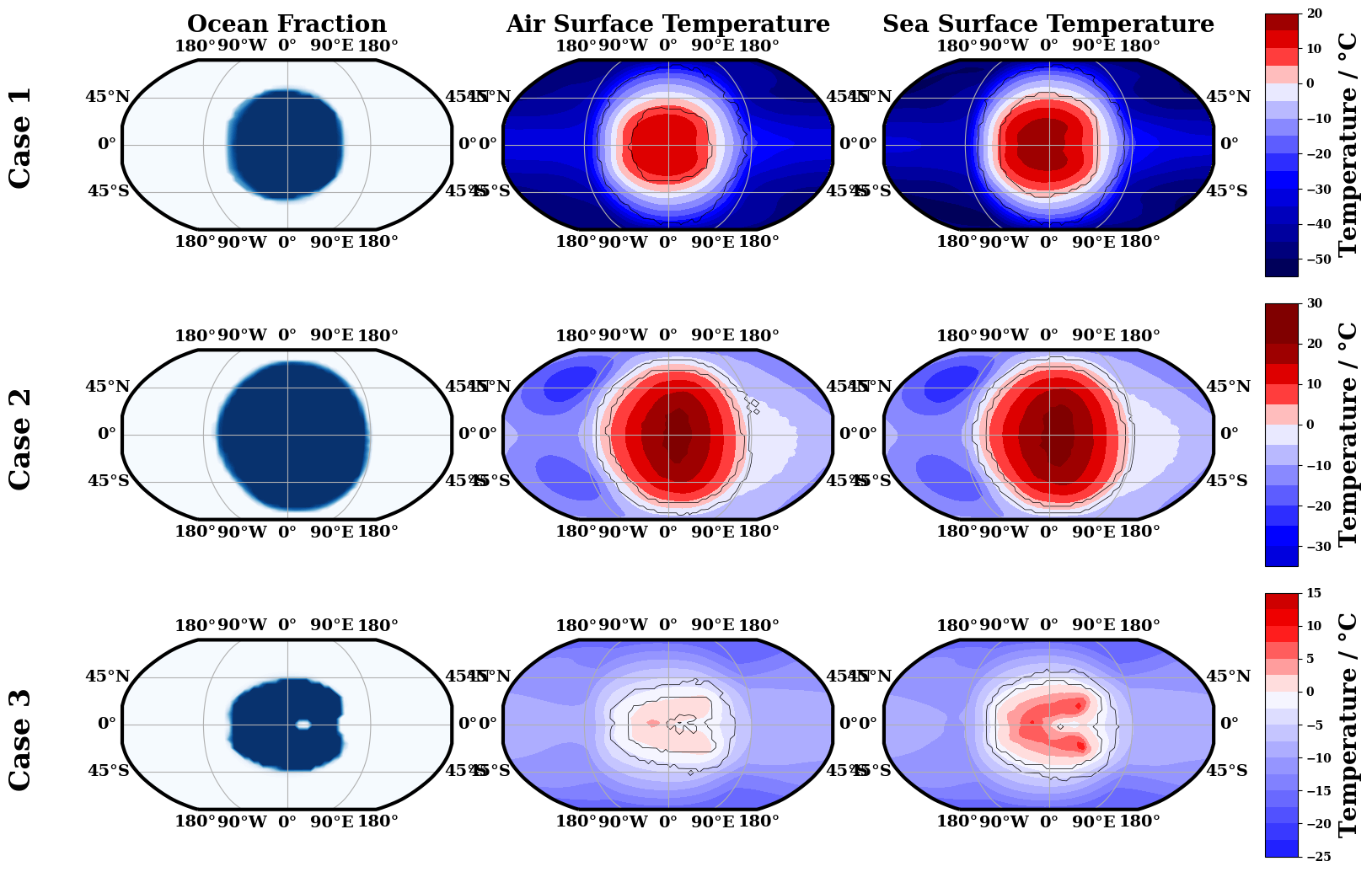}
    \caption{Left: Average ocean fraction, calculated assuming a seawater freezing point of $-1.8\degree C$ for Cases 1 and 3 and $0\degree C$ for Case 2. White shows the ice, and blue ocean. Middle: The mean near-surface air temperatures. Shown within the black contours are the regions where these vary between above and below freezing over the course of a single Earth year. Right: The mean surface temperatures. Also shown within the black contours are the regions where these vary between above and below freezing over the course of a single Earth year: these regions are narrower than the air `perifreezing zone'. }
    \label{fig:perifreezings}
\end{figure*}

Next, we focus on the presence of sea ice, which forms when the surface temperatures are below the freezing point of seawater: $-1.8\degree C$ for Cases 1, and 3 and $0\degree C$ for Case 2. Figure \ref{fig:perifreezings} shows the time-averaged open ocean fraction, and the air and surface temperatures of our various cases. The `air temperature' is the temperature of the lowest atmosphere layer (i.e. the atmosphere boundary layer, which is usually the bottom few hundred metres of the atmosphere), whilst the surface temperature is the temperature of the slab ocean or of the sea ice top. 

In both instances we also plot what we call the `perifreezing zone', defined as the area where temperatures reach both above and below freezing ($-1.8\degree C$ or $0\degree C$ depending on the ocean salinity). This is diagnosed by taking instantaneous readings once every Earth day for the final Earth year of the simulations, and allows us to probe time variability on medium timescales.  This `perifreezing zone' is important as it may indicates zones where ice formation and melting can happen, as we discuss in Section \ref{sec:seaice}. The `air perifreezing zone' represents areas where we may see some small scale melting or freezing. The `surface perifreezing zone', representing zones where the average temperature of whole grid cells fluctuates above and below freezing, indicates areas with large scale freezing and melting.

Affected by the exact instellation and planetary parameters, the exact amount of ice cover varies between the simulations. However, in all cases there is dayside ice present as well as open ocean, so an ice-water interface and a zone where temperatures fluctuate around freezing. 
In Case 1, which has a thin atmosphere without large sources of opacity but a relatively high instellation (900 Wm$^{-2}$), there is liquid surface water between roughly 60\degree W and 60\degree E, and 45\degree N and 45\degree S. The `air perifreezing zone' is quite wide -- several grid cells -- which is a reflection of the atmospheric variability on TRAPPIST-1 e, which can take the form of an eastwards-travelling planetary-scale wave \citep{Sergeev2022}. In contrast, the sea temperature is buffered by the significant heat capacity of the wind-mixed layer and so varies less. This means that only the grid cells with mean temperatures close to freezing point have surface temperatures that fluctuate above and below freezing, so the `surface perifreezing zone' is only a few grid cells wide. However, it is important to note that the grid cells are very large, 4\degree \, latitude by 5\degree \, longitude, corresponding to 400 by 500~km for Case 1, and so that even this relatively thin surface perifreezing zone is well over a thousand km across.

In Case 2, the reduced instellation is more than compensated by the increased thickness of the 5 bar CO$_2$ atmosphere, and there is liquid water on the vast majority of the dayside, brushing the south-eastern terminator. There is a slight North-South asymmetry in the ice cover and surface temperature maps, which is puzzling: there are no signs of any asymmetry in the surface or zonal-mean zonal winds that might explain such a difference in heat transport, and it is possible that this is simply a result of model stochasticity. The thickness of the CO$_2$ atmosphere reduces the horizontal but also temporal temperature variations. Net, this leads to a thinner air perifreezing zone, and a surface perifreezing one to two grid cells across. Changing the ocean salinity and so its freezing point does not seem to have much of an influence: whilst it leads to a slightly smaller open fraction and slightly shifts the location of the perifreezing zone, having ice form and melt as one temperature instead of two slightly different ones does not seem to have a significant impact. We think this is \textbf{because} the temperature of the ice bottom is set at the ocean freezing temperature: if the temperature of the ice interior rises above this, the ocean-ice heat flux is quite high and will rapidly melt the ice.

Case 3, in contrast, has the greatest amount of ice cover. There is open ocean between about 45\degree W and 45\degree E, and 30\degree N and 30\degree S. Despite the higher instellation, the anti-greenhouse effect of methane and the enhanced Rayleigh scattering combine to efficiently reduce surface temperatures. There is one intriguing feature, and that is an isolated patch of ice, just east of the substellar point. We have investigated this region: it emerges from the increased cloud cover and precipitation around the substellar point. It appears that, due partly to the increased cloud cover and decreased shortwave surface flux, this region has average surface temperatures above the freezing point of seawater ($-1.8\degree C$), but below the melting point of ice ($0\degree C$). There is thus a constant heat flux up from the ice bottom leading to ice melting at the bottom, as discussed above. However, there are significant amounts of precipitation falling as snow on the top of the ice (an order of magnitude more than in the rest of the perifreezing zone). As the surface is below freezing this is converted to ice, maintaining the thickness of the sea ice here. We note that ice patch is also a consequence of the slab ocean model not allowing for ocean heat transport, and that a more realistic model would be unlikely to show this same ice distribution. The Case 3 air temperature variations are reasonably large, although not as large as in Case 1, while the surface perifreezing zone -- though now oddly doughnut shaped, following the ice cover -- is again only a one to two grid cells across in most places, except for the ocean eastern edge. This is, in part, wider due to the variation in cloud coverage: as can be seen in Figure \ref{fig:innpri_gridcell}, the incident shortwave irradiation varies heavily. In the zone east of the substellar point, it varies by a factor of two, driving temperature variations over a larger scale. However, it is also clear that there are temperature variations on a longer timescale than the cloud variability, so some other feature of the atmospheric dynamics -- maybe a longer-term oscillation or wave-mean flow interaction -- must also be promoting time variability there.

\subsection{Sea Ice Thickness, Freezing and Melting}

The presence or not of sea ice is only one metric. Other useful properties include the sea ice thickness and the average ice melting and formation rates. The proportion of ice in a given location which is "first year ice" (ice less than a year old) is a useful measure of ice turnover, and another useful high-level metric. These quantities are key characteristics of the ice, and are important for assessing its stability and potential to concentrate nutrients. It is also a key parameter for assessing the possibility of retaining impactor fragments, as discussed in Section \ref{sec:seaice}.

Presenting global maps of these quantities is not the most effective method, as these quantities are usually zero outside a small strip on the ice shelf edge. Additionally, as the sea ice surface on the nightside is permanently freezing and receives negligble latent heat flux from precipitation and no stellar irradiation, the nightside ice thickness unphysically grows steadily over time. The only balance to this is the internal heat in our slab ocean parametrisation, which leads to a slight ocean-ice heat flux. However, in steady state with a freezing surface, this would take thousands of model years to reach a steady state with an ice sheet several hundred meters thick (evaluated using our ice model described in Section \ref{sec:seaice_icemargins}). Our GCM simulations do not begin to approach this regime, and we thus end up with unphysical monotonic ice growth. This, however, is not an issue in the perifreezing zone, where the variations in surface and ocean forcing generally constrain the ice thickness to appropriate levels, even with a general underestimation of the ocean heat flux. 

\begin{figure*}
    \centering
    \includegraphics[width=0.95\textwidth]{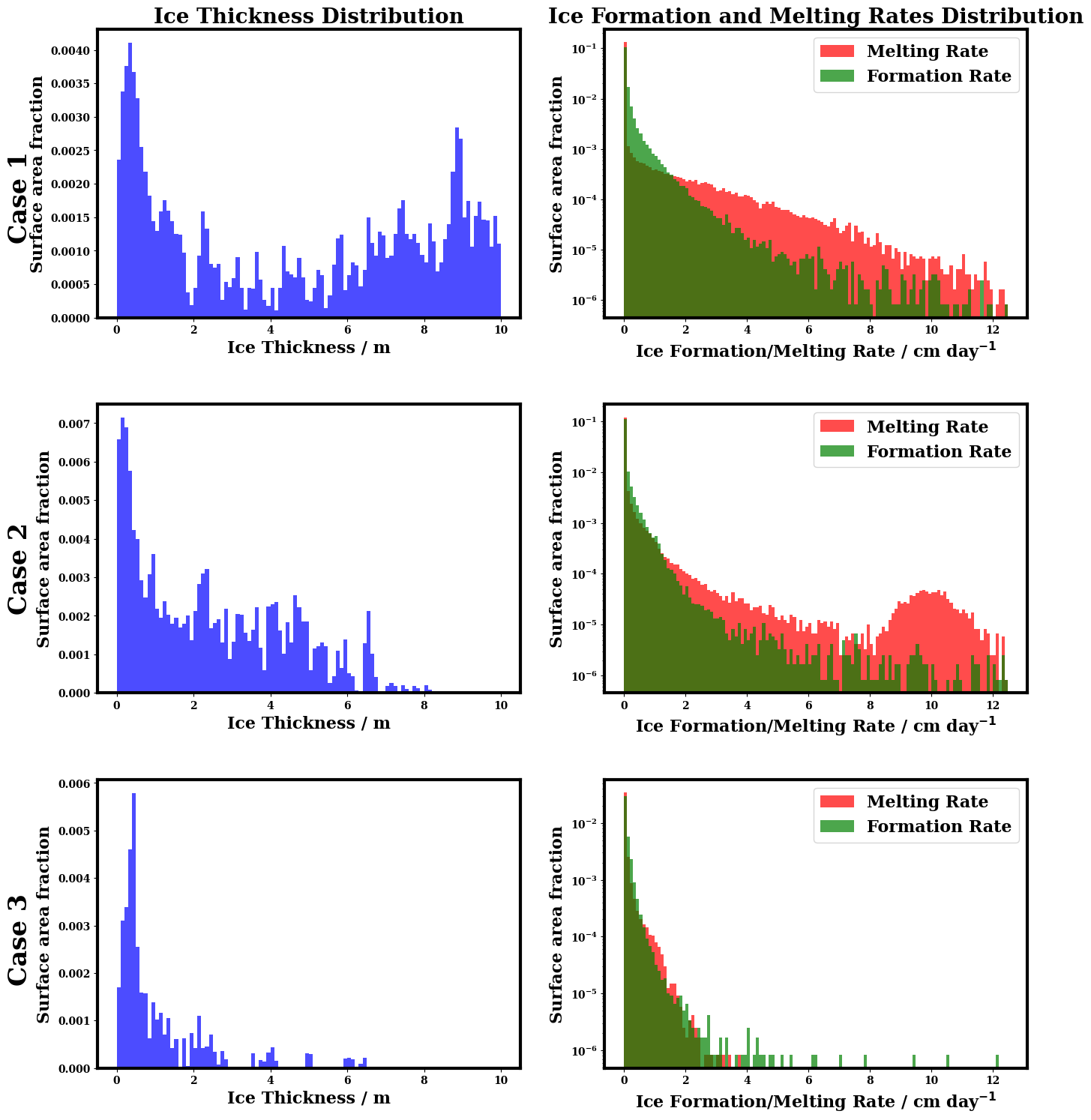}
    \caption{Left: The distribution of ice thickness in the surface perifreezing zone, taken from daily data in the last model year of simulation. Right: The distribution of ice melting and formation rates in the surface perifreezing zone. There is a large amount of ice less than a few metres thick. The tail of higher ice thicknesses, most present in Case 1, is a consequence of ocean heat transport not being included in the model: the surface perifreezing zone includes some regions which are almost always freezing and so have grown thick ice. At each time step, the majority of the ice in the perifreezing zone is neither melting nor growing. However, there is significant intermittent freezing and melting freezing. These rates commonly reach up to 10 cm/day for Cases 1 and 2 and 2cm /day for Case 3, leading to the complete melting or creation of thick ice in a short timeframe.}
    \label{fig:ice_props}
\end{figure*}

Instead, in Figure~\ref{fig:ice_props}, we show histograms of three key properties in the surface perifreezing zone: the ice thickness distribution and the ice melting and growth rates, presented as fractions of the planet's surface they cover. 
We additionally show the average fraction of the planet's surface covered in first year ice in Table \ref{tbl:fyice}.

\begin{table}
\centering
\begin{tabular}{|c|c|}

\hline \hline
\textbf{Case Name} & \textbf{First year ice fraction} \\
\hline
Case 1 & 0.0155 \\ 
Case 2 & 0.0210 \\
Case 3 & 0.0030\\ 
\hline \hline
\end{tabular}
\caption[First Year Ice Fraction]{Fraction of the planet's surface covered in ice younger than a year for our different simulations. This is a small fraction but still corresponds to millions of km$^2$ in all cases.}\label{tbl:fyice}
\end{table}

The perifreezing zone makes up a substantial fraction of the planet's surface (2-10\%). Much of this region has low ice thickness ($<1$m), with the median thickness at $\sim$0.5~m. There is then a tail of ice thickness distribution up to about 10~m thick, particularly in our Case 1 simulation. The thicker end of this tail is likely an artefact of our lack of a dynamic ocean: in regions which are only very rarely above freezing, the lack of ocean heat transport favours the ice steadily thickening over long enough timescales; the inclusion of a dynamic ocean would likely also change the shape of the perifreezing zone \citep[e.g.][]{Hu2014}. We do, however, also note that the lack of ice drift means that ice ridging and dynamical ice thickening more generally is not accounted in our model. This would likely produce an increase in ice thicknesses in our perifreezing zones, which generally have convergent winds.

Whilst most of the perifreezing zone is not melting or freezing at any given moment, the whole perifreezing zone does undergo melting and freezing, as we have verified through animated maps of the melting and growth rates. This confirms the surface perifreezing zone as a reasonably good metric of areas of large-scale ice formation and destruction. The ice formation and melting rates seem to follow, roughly speaking, a power-law distribution, and are mostly below 2 cm/day for Cases 1 and 2 and 1 cm/day for Case 3. In Cases 1 and 2, they rarely exceed 10 cm/day, and in Case 3, they very rarely exceed 2 cm/day, reflecting its the lower degree of time variability.
These are large scale averages: sub-areas of each of the grid cells will have rates which are higher than this, and contribute to a significant amount of ice turnover. Overall, 0.3\% to 2\% of the surfaces are covered by relatively new `first year' ice in our simulations. This is a minority of the perifreezing zone and a small fraction of the planet's surface, but we note that it still corresponds to millions of km$^2$ in all our cases. In absolute terms, this is still a vast area and so there are large amounts of ice formation and destruction.

\subsection{Climate Variations in the Perifreezing zone}

\begin{figure*}
    \centering
    \includegraphics[width=0.95\textwidth]{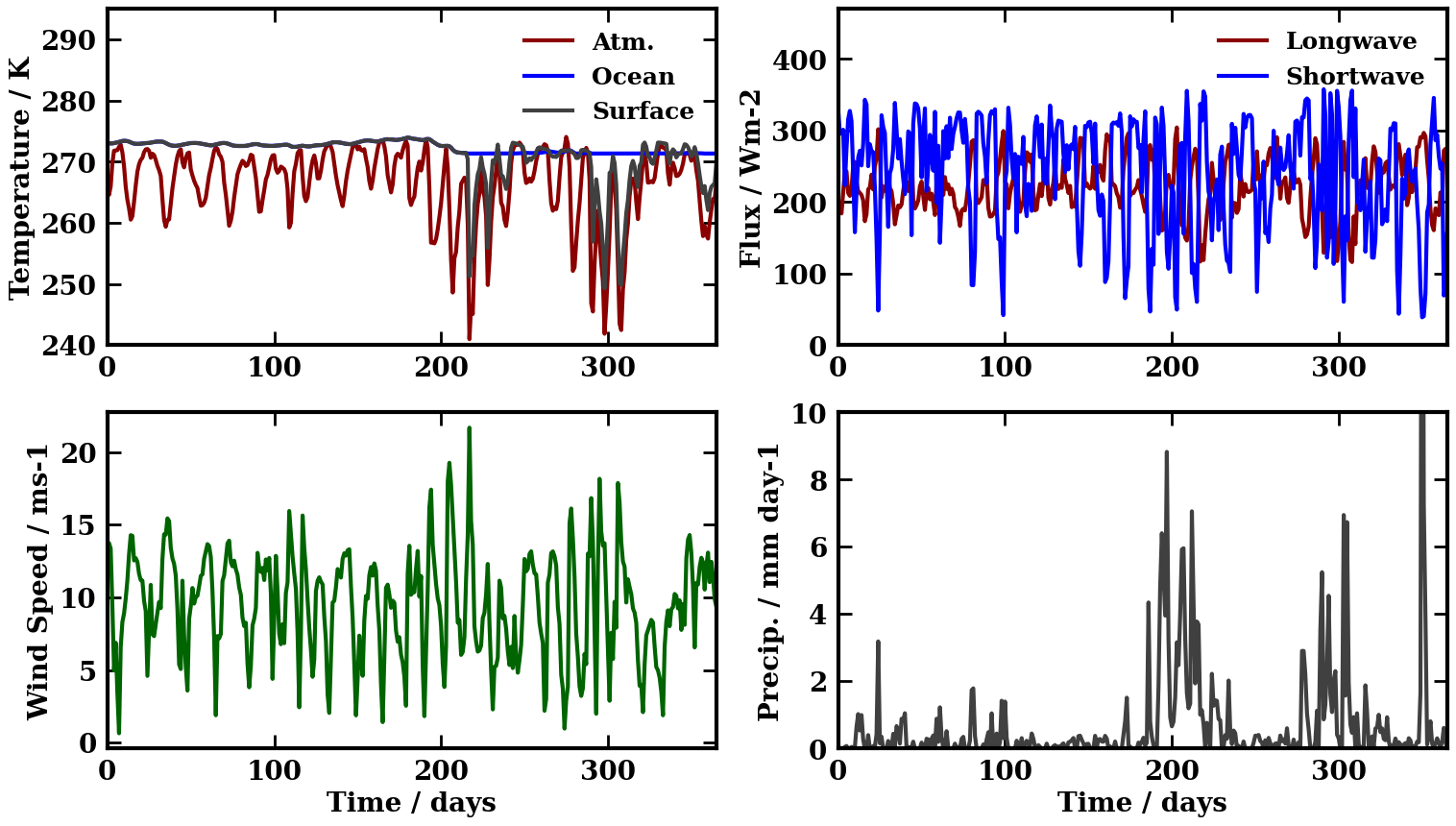}
    \caption[Case 1 grid cell]{The grid-average climate conditions in the Case 1 grid cell centred at 46\degree$ S$, 37\degree $ W$. Top left: surface, ocean, and atmosphere temperatures. Top right: Longwave and Shortwave radiation fluxes incident at the surface. Bottom left: surface wind speed. Bottom right: total precipitation (ice + liquid). There are significant periodic atmospheric temperature variations of 10-30 K, damped at the surface. Longwave and shortwave incident fluxes are about the same magnitude. Wind variations are associated with the temperature fluctuations, and precipitation is highest when the temperature fluctuations are largest.}
    \label{fig:inn2nd_gridcell}
\end{figure*}

\begin{figure*}
    \centering
    \includegraphics[width=0.95\textwidth]{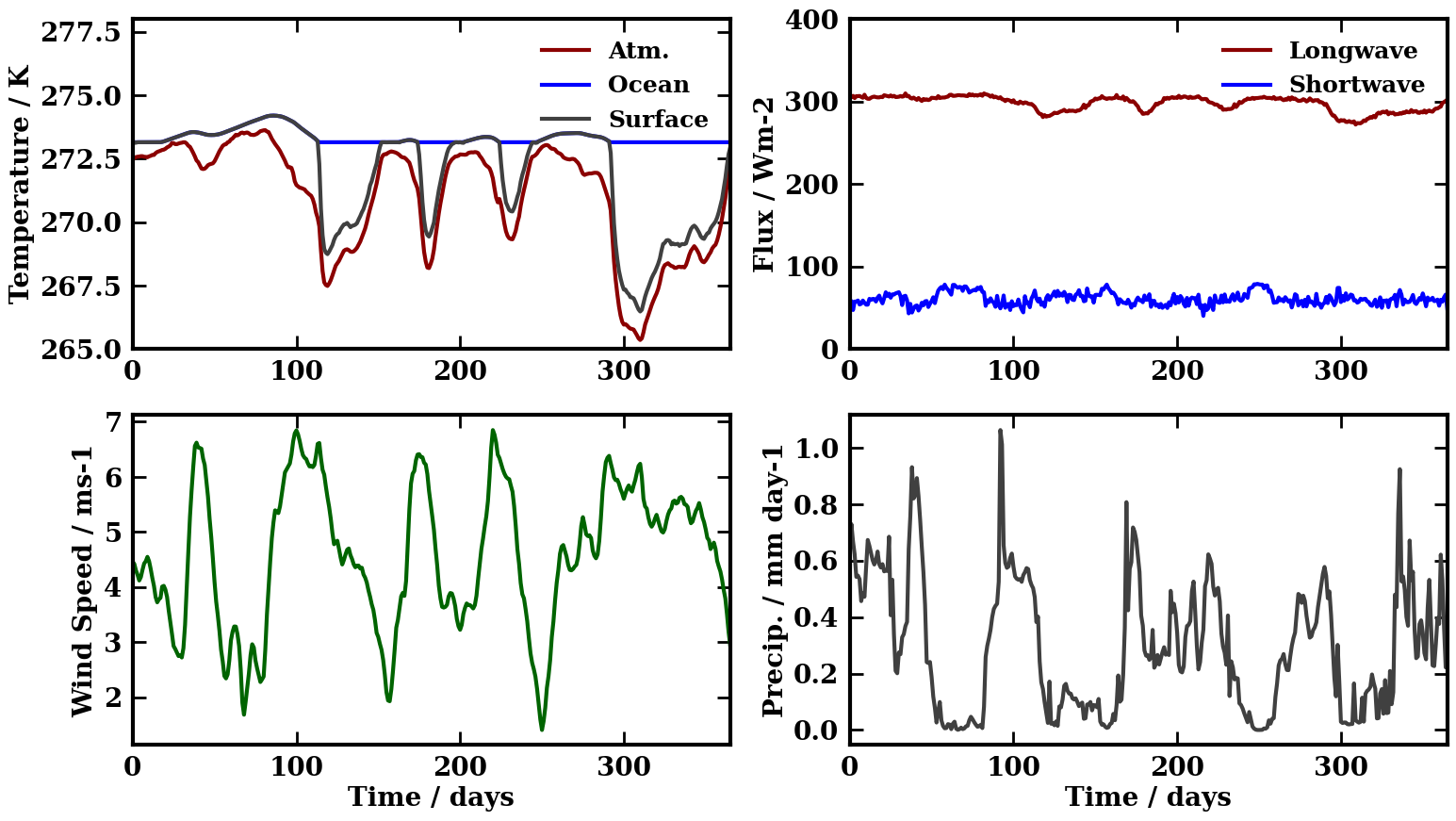}
    \caption[Case 2 grid cell]{The grid-average climate conditions in the Case 2 grid cell centred at 74\degree$ S$, 8\degree $ W$. Top left: surface, ocean, and atmosphere temperatures. Top right: Longwave and Shortwave radiation fluxes incident at the surface. Bottom left: surface wind speed. Bottom right: total precipitation (ice + liquid). The surface alternates between frozen and melted on timescales of about 50 days, and wind and precipitation also vary on these timescales. Given the grid cell is close to the terminators, the incident shortwave radiation is much lower than the longwave.}
    \label{fig:out2nd_gridcell}
\end{figure*}

\begin{figure*}
    \centering
    \includegraphics[width=0.95\textwidth]{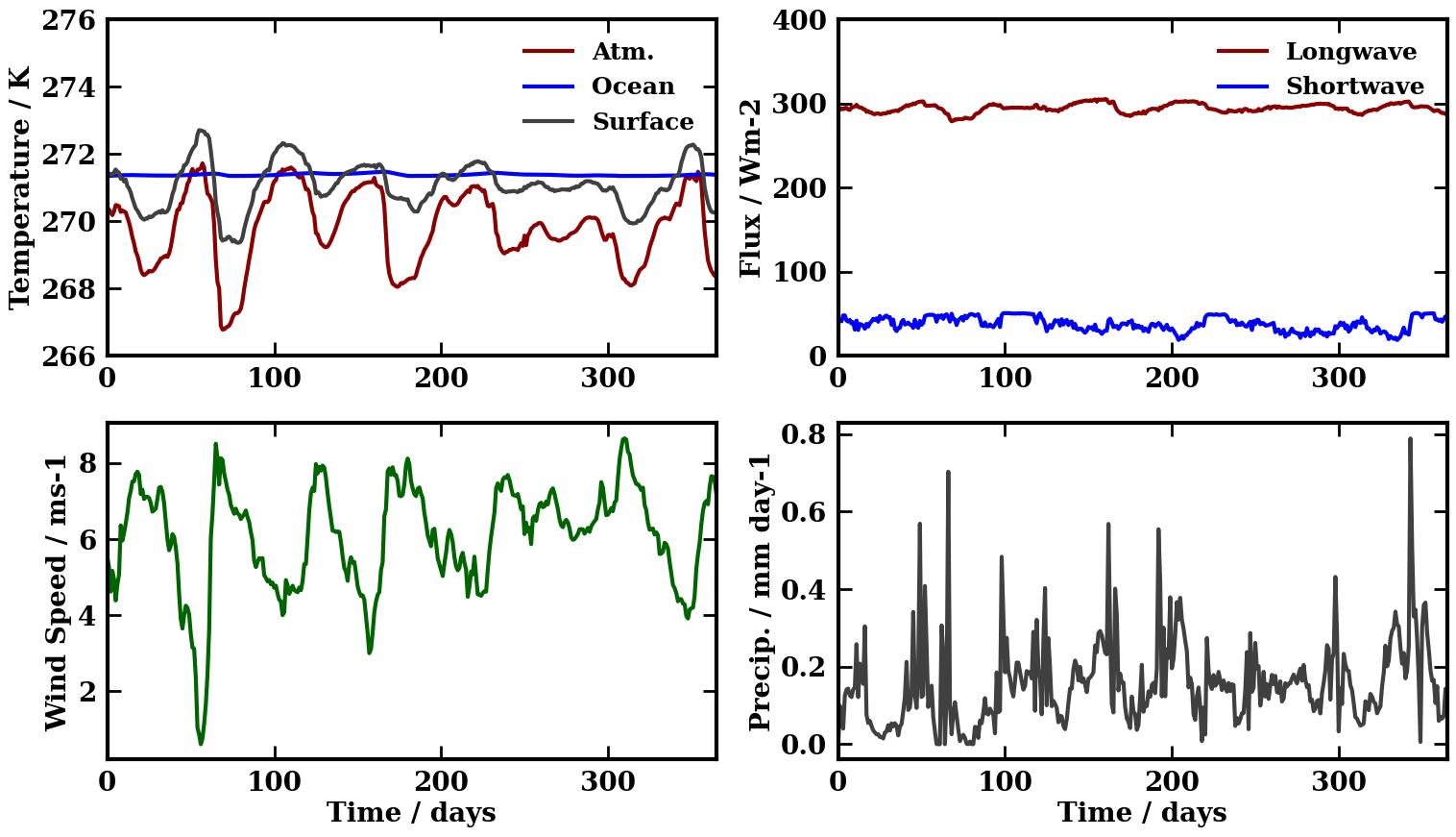}
    \caption[Case 3 grid cell]{The grid-average climate conditions in the Case 3 grid cell centred at 34\degree$ S$, 53\degree $ E$. Top left: surface, ocean, and atmosphere temperatures. Top right: Longwave and Shortwave radiation fluxes incident at the surface. Bottom left: surface wind speed. Bottom right: total precipitation (ice + liquid). In this cell, the ice fraction only ever dips slightly below one, even as the surface is occasionally warmer than the ocean. Temperatures and winds vary on 30-50 day timescales, and the high optical depth of the atmosphere means that there is little incident shortwave radiation.}
    \label{fig:innpri_gridcell}
\end{figure*}

Whilst it is useful to study the variations above and below freezing and the growth and melting rates of the ice, it is also informative to consider more precisely the temperature variations present in and around the peri-freezing zone. Figures \ref{fig:inn2nd_gridcell}, \ref{fig:out2nd_gridcell}, and \ref{fig:innpri_gridcell} show a year-long timeseries of the grid-average climate conditions in three different grid cells across our three cases, all located in the surface perifreezing zone. We show the surface, ocean, and atmosphere bottom temperatures, along with the surface longwave and shortwave fluxes, the average surface winds ($\sqrt{u^2+v^2}$) and the precipitation rate, giving us a detailed view of factors affecting the sea ice.

Some features immediately stand out. Case 1 has an atmospheric perturbation travelling east that a period of about 15 days or 2.5 orbits, which matches some of the atmospheric variability in \citep{Sergeev2022}. This leads to significant atmospheric temperature variability with an amplitude of about 10~K, but sometimes as high as 30~K. The response of the surface temperature to this forcing varies: it is noticeable that when the ocean is frozen over ($\approx 210-240$ and $290-320$ days in Figure \ref{fig:inn2nd_gridcell}), the surface temperatures show much greater variation. This is because the top of an ice sheet is more strongly coupled to the atmosphere than the 100m thick slab ocean, which has a much higher thermal inertia. In contrast, Cases 2 and 3 (with thicker and more opaque atmospheres) show more infrequent and smaller fluctuations, with fluctuations of up to 5~K over timescales of a few months. In Figure \ref{fig:out2nd_gridcell}, the temperature and radiation forcing leads to alternating ice cover and open ocean, giving an example of large-scale freezing and melting occurring and how climate variability contributes to this. 

We have chosen the grid cell in Figure \ref{fig:innpri_gridcell} to highlight a slightly different regime. Here, the long-term climate average is cold enough to maintain permanent near-complete ice cover (there are some small variations in the average ocean temperature too small to make out in the plot, and the fractional ice cover sometimes dips below 1). However, the surface temperature does occasionally rise above the ocean's freezing temperature, which results in a degree of ice bottom melting, and then freezing when temperatures drop again. So even in locations where the ice cover is permanent, there are still temperature fluctuations around freezing, and some large-scale melting and freezing. 

A few other climate features can be highlighted in these figures. The higher shortwave flux on Case 1 (TRAPPIST-1 e analog) can be seen, as can the variability caused by clouds. The increased activity of the water cycle can also be seen in the precipitation rates, which are almost an order of magnitude higher for Case 1 than Cases 2 and 3. The lower shortwave flux in Case 2 can be attributed to the perifreezing zone, including the grid cell chosen, being close to the terminator, illustrating how factors such as the precise location of the perifreezing zone can affect prebiotic chemistry potential and planetary habitability. Finally, the wind variability, which is clearly correlated with the temperature variations, is apparent in all cases. This will affect the magnitude and direction of sea ice drift.

\subsection{Winds}

\begin{figure}
    \centering
    \includegraphics[width=0.48\textwidth]{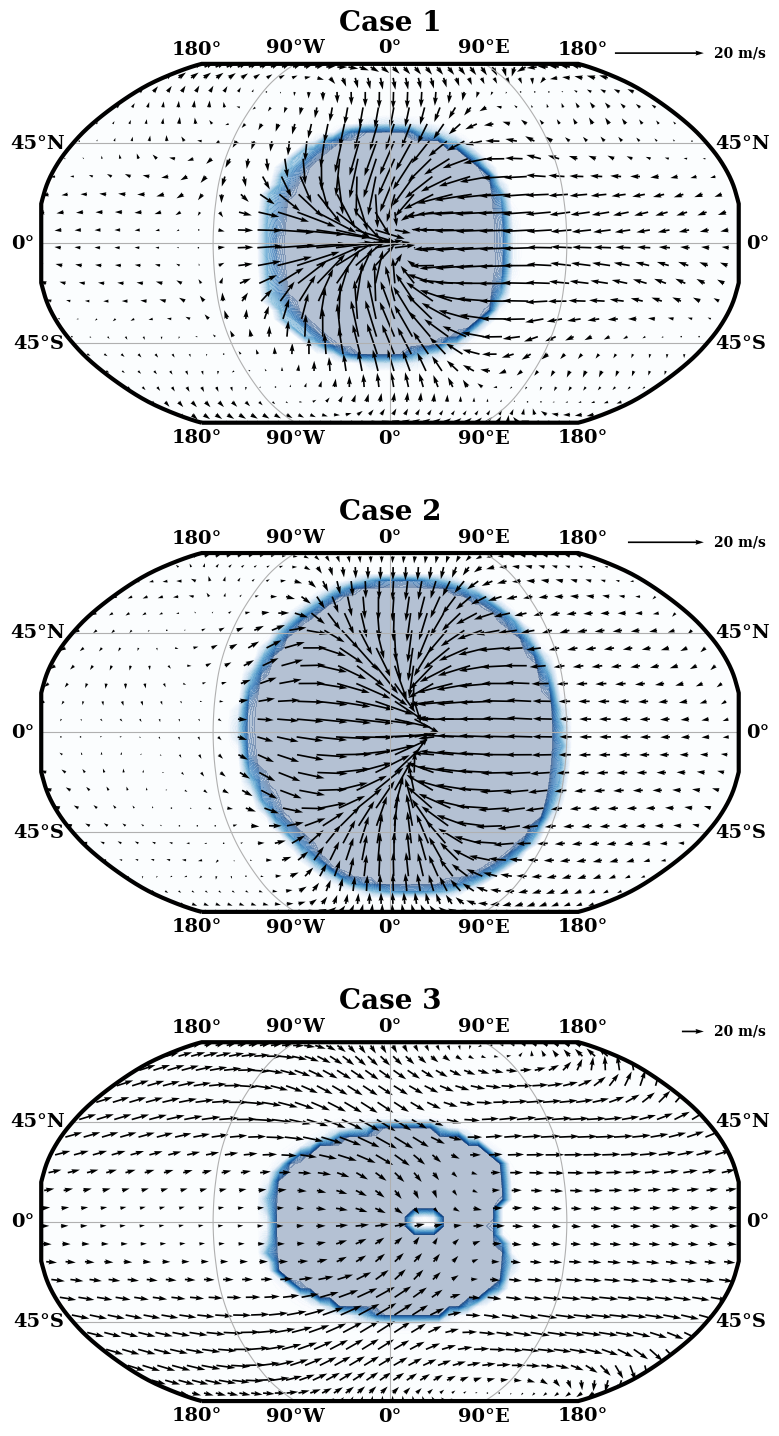}
    \caption{The mean surface winds for our different cases, over-plotted with the average ocean fraction (blue: ocean, white:ice). We usually see a convergent motion around the substellar zone as part of the general dayside-nightside overturning circulation. However, the details of the atmospheric dynamics vary between each case and there are normally locations at the ice edge where the prevailing winds run perpendicular to the ice edge or even into the ice sheets.}
    \label{fig:surfwinds}
\end{figure}

Another important set of climate features are the surface winds, whose time averages are shown in Figure \ref{fig:surfwinds}. As we will discuss in more detail in Section \ref{sec:seaice}, winds create a drag force on the sea ice and can cause it to drift. The full impact of this is not something that is represented explicitly by the model: first, because our GCM does not include large-scale sea ice drift \citep[unlike, e.g.][]{Yang2020}, which might cause changes in the ice distribution, as discussed in Section \ref{sec:summary}. Secondly, the large length-scales examined in GCMs prevent an examination of small-scale ice drift and behaviour. Here, our grid cells have widths of $\mathcal{O}(1000km)$: while GCMs with adaptive resolution can have much higher local resolutions \citep[such as 4.7km in ][]{Sergeev2024} and so provide more detailed simulations of sea ice growth and drift, their resolutions are still slightly too coarse to simulate the details of ice packs, leads, and openings, which have spatial scales of $\mathcal{O}(100m-1km)$ \citep{Thomas2010}.

The average winds in the vicinity of the ice shelf edge vary widely between simulations. In general, the tidally-locked cases show a convergent motion towards the substellar point. This is part of the general dayside-nightside overturning circulation, and is associated with divergent movement at higher altitude. One might expect that this would lead to universal ice drift towards the substellar point, perpendicular to the ice-open ocean boundary. This is indeed largely the case, but the complexities of the atmospheric circulation mean that this is far from the only option. Whilst we do not conduct a full analysis of the atmospheric circulations, overviews of similar cases can be found in e.g. \citet{Sergeev2022} for the Case 1 and in \citet{Barrier2025b} for Case 3.

In Case 1, the presence of standing extra-tropical Rossby waves \citep[e.g.][]{Carone2015} can still be seen at the surface as four weak high-latitude gyres: two clockwise in the western hemisphere and two anticlockwise in the eastern hemisphere. If the instellation were a higher and the open ocean zone somewhat wider, the prevailing wind directions would be perpendicular to or even into the ice sheet, although the increased instellation might also affect the dynamical circulation and the average surface winds. In Case 2, there is no equivalent gyre impact and the circulation is largely convergent on the dayside. However, the wind motion around the substellar point is not exactly convergent: for both Cases 1 and 2, at points on the southwestern and northwestern edge the average wind is nearly perpendicular to the ice edge. 

In Case 3, we can again see the extra-tropical Rossby wave, but the magnitude of the eddy-rotational and divergent wind components are weaker compared to the jet rotational circulation, which leads to eastwards winds across most the surface. This means that the dominant winds encourage ice drift into the open ocean on the western region of the substellar zone, but into the ice sheets on the eastern edge, and largely perpendicular to it at the north- and south-east corners. The ice patch just east of the substellar point further increases the proportion of the ice edge where the dominant winds go towards the ice shelves, although we re-emphasize that the ice presence here would likely not be replicated in a GCM with a dynamic ocean.

\begin{figure}
    \centering
    \includegraphics[width=0.48\textwidth]{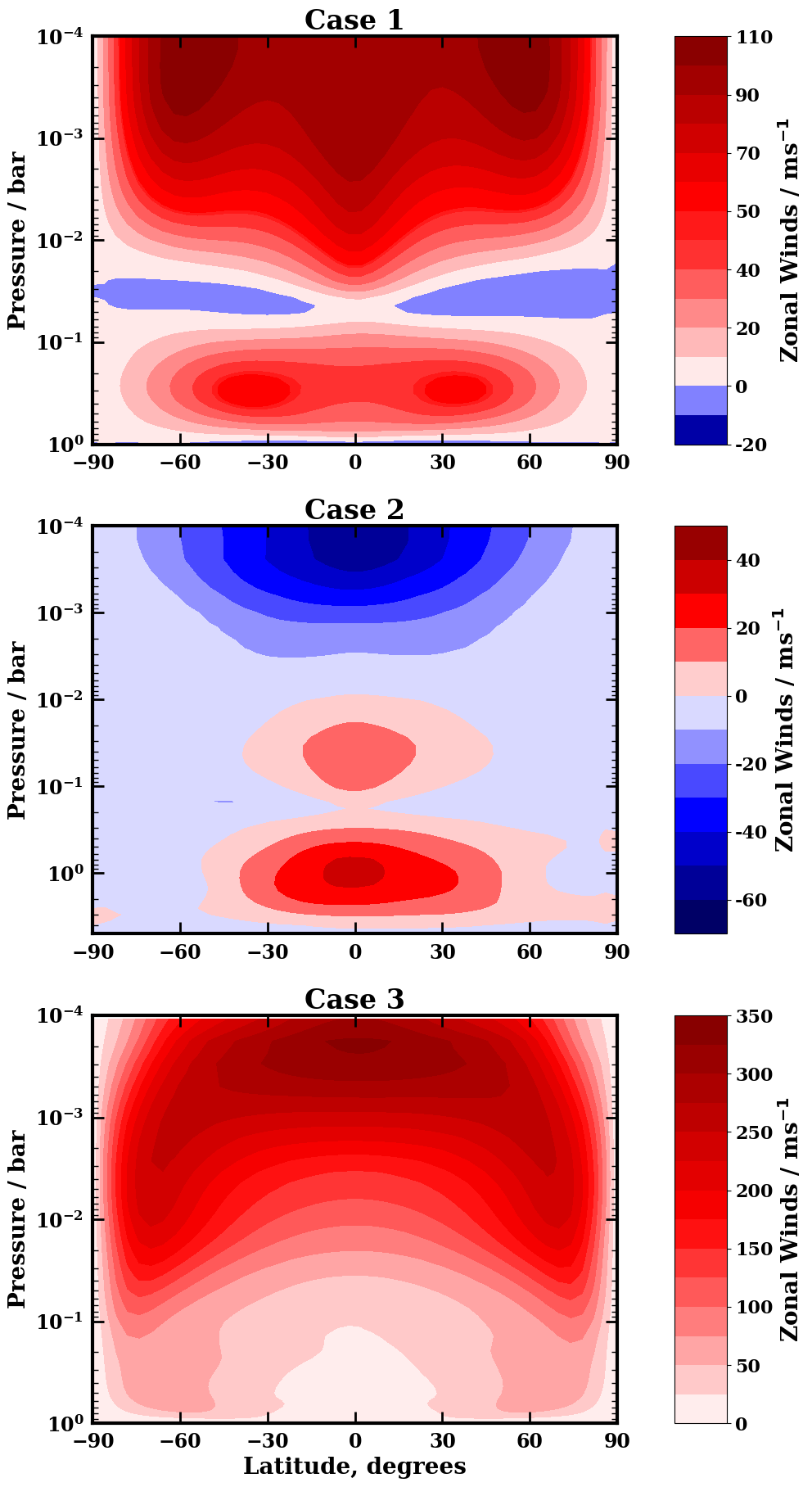}
    \caption{The zonal mean zonal winds for our different cases. All show tropospheric equatorial super-rotation, although in Case 3 the \textbf{peak} axial angular momentum latitude is found at higher latitudes in the form of high-latitude zonal jets. Cases 1 and 3 also show broad stratospheric super-rotation, but in Case 2 there is a retrograde jet towards the model top.}
    \label{fig:u_zonal}
\end{figure}

We also show an aspect of the overall atmospheric circulation in Figure \ref{fig:u_zonal}, where we show the zonal-mean zonal winds. The zonal winds control much of the dayside - nightside heat redistribution and their advection can strongly affect the cloud distribution.  In all cases, we largely find tropospheric super-rotation. In Cases 1 and 2 this is takes the form of a super-rotating equatorial jet (relatively broad for the former case and narrow for the latter), but in Case 3 the axial angular momentum peaks towards the mid-latitudes, as seen in some of the simulations in \citet{Barrier2025b}.  
In Cases 1 and 3 there is strong and broad stratospheric super-rotation, although in Case 2 (representing LHS 1140 b) we find a stratospheric westwards equatorial jet towards our model top \citep[resembling, for example, the 100x metallicity case in][]{Charnay2021}, which we think may be due to the model top or wave - mean flow interactions. We also note that at these altitudes the circulation is dominantly dayside-nightside overturning, and so an apparent retrograde jet actually means that the western hemisphere overturning circulation is faster than the eastern hemisphere one. In all cases, the eastwards tropospheric winds contribute to the clouds being advected eastwards, as is seen in Figure \ref{fig:cld_fsds}, with the higher latitude cloud bands in Cases 1 and 3 possibly linked to the higher latitude superotation winds there.

\subsection{Clouds and Radiation}

\begin{figure*}
    \centering
    \includegraphics[width=0.95\textwidth]{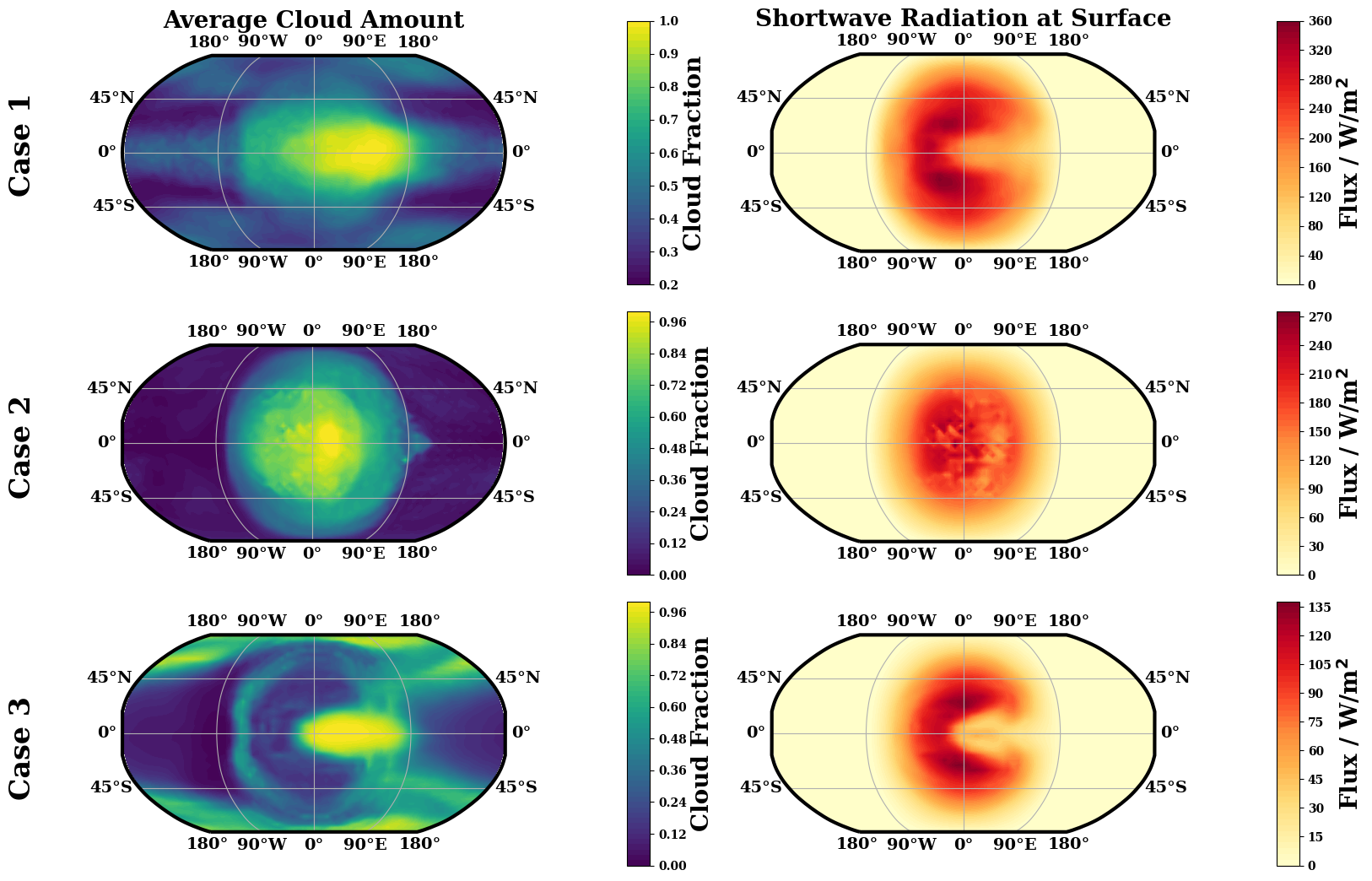}
    \caption{The mean cloud cover and downwards shortwave surface radiation for our different cases. Clouds are normally more concentrated on the dayside, but the exact distribution of these varies, and is affected by the precise climate structure and atmospheric dynamics: they can also be common in colder nightside gyres. The downwards shortwave radiation varies greatly between cases. It is high in thinner, lower optical depth atmospheres, but much lower in H$_2$ dominated cases with enhanced scattering.}
    \label{fig:cld_fsds}
\end{figure*}

The final set of relevant climate features are the cloud patterns and the shortwave radiation incident at the surface. These are shown in Figure \ref{fig:cld_fsds}, and are very relevant in assessing the potential for abiogenesis. For prebiotic chemistry to occur, it is not enough that nutrients are concentrated: an energy source is also needed. Here, stellar radiation provides a free energy source for prebiotic chemistry, and UV radiation in particular can catalyse certain reactions. There is generally a substantial cloud cover on the dayside which reduces the incident stellar radiation. This cloud cover can be unevenly concentrated and still leaves a significant fraction of shortwave radiation reaching the surface. 

In Case 1, dayside clouds are concentrated in a thick equatorial band, with lower cloud cover at higher latitudes. This is reflective of the higher temperatures and greater humidity in the substellar regions, as well as the general overturning circulation leading to rising air on the dayside. Incident stellar radiation is thus highest in the dayside mid-latitudes, reaching $\sim 350$ Wm$^{-2}$, about the same as the planet-averaged top-of-atmosphere flux on Earth. In Case 2, the relatively weaker equatorial jet and the stronger influence of the isotropic day-night overturning circulation leads to thick clouds on the whole dayside, concentrated towards the substellar point. Despite the thick CO$_2$ atmosphere, large regions of the surface have an incident flux of around 200 Wm$^{-2}$. 

The dayside clouds in Case 3 are relatively weak, except at the substellar point and east of it. However, the H$_2$-dominated atmosphere with significant additional opacity from CH$_4$ and the enhanced Rayleigh scattering contribute to blocking about 90\% of the incident radiation from reaching the surface, with the highest surface shortwave flux $\sim 140$ Wm$^{-2}$. This is lower than the other cases and a reflection of the high opacity of H$_2$-dominated atmospheres. Prebiotic chemistry might thus be somewhat hindered in such atmospheres, especially if they are thicker than the 1 bar atmosphere presented here. This would also likely also have a negative impact on the biological productivity of any biosphere present, assuming that they are phototrophs.

\section{Nutrient concentration on ice shelves}

\label{sec:seaice}

The mere presence of dayside sea ice, by itself, does not necessarily mean that it is a plausible location for prebiotic chemistry to lead to abiogenesis. Bioessential molecule concentration needs to happen, followed by the right conditions for prebiotic chemistry. In this section we start by drawing on our GCM simulations to discuss more generally what sea ice behaviour we might expect on exoplanets with deep oceans. We then investigate how concentration of feedstock molecules may emerge in two main ways: either from sea ice formation at the edge of ice shelves, or alternatively by impactor fragments landing on the ice. 

\subsection{Ice shelf properties on exoplanets}

In the introduction, we summarised some key aspects of sea ice formation and behaviour on Earth. We can now start to consider how the various aspects of the planetary climate seen in Section \ref{sec:gcm} will affect the sea ice, moving beyond some of the specific results seen to a more general prediction for sea ice on tidally-locked planets. In general, we think that the paradigm of a permanent ice sheet and a perifreezing zone with climate fluctuations driving large scale freezing and melting, as we find in Section \ref{sec:gcm}, is a robust one we expect to be common across planets with sea ice.

An immediate difference from Earth is that the simulations we perform, having deep global oceans, are tidally locked and continent-free. Changing the position of number of continents can have significant effects on the climate \citep{Lewis2018,Macdonald2022}, but in our cases we expect a more symmetric (especially North-South) climate and atmospheric circulation, which would correspondingly influence the shape of the ice-ocean boundary.

For tidally locked planets with no obliquity, the ice distribution would be in roughly steady state (as shown in Figures \ref{fig:perifreezings} and \ref{fig:ice_props}). The inclusion of realistic ocean circulation and ice drift would change our results somewhat, by affecting the ocean-ice heat flux and by changing the surface albedo respectively.
We expect a smooth gradient of average ice thickness controlled by the local air and ocean surface temperatures (which dictate the ocean-atmosphere heat flux). Right at the ice edge, the different planet atmospheres we simulate have average ice thicknesses of 0.1-1 m. Ice thicknesses are expected to be moderate globally: they are generally under 5 m in \citet{Yang2020}, with a fairly smooth change in average ice thickness moving from the cooler to the hotter regions of the planet. The shape of open ocean zone would be changed by the ocean circulation \citep[e.g.][]{Hu2014}, and sea ice drift can lead to additional complexity in the shape and extent of the open ocean zone \citep{Yang2020}. 

We think that large scale ice melting and freezing will happen in the region we call the `perifreezing zone', which will -- depending on climate variability -- vary from dozens to thousands of km across. This zone is where grid cell average surface temperatures vary above and below the freezing point of seawater, and our simulations confirms ice growth and destruction is concentrated there. However, smaller-scale ice formation and melting could happen outside this zone. For example, a solar beam may be incident on a suitably shaped section of ice. This may form a small melt pool, whose lower albedo leads to continued melting there, and there might be some persistent liquid water in otherwise freezing conditions. Alternatively, in partially cloudy regions there will be cloud-free regions where the incident stellar radiation is higher than the grid cell average (which is calculated using the average grid cloud fraction), and the full stellar beam is incident on the ice. This could increase the local ice temperature above freezing, forming a small melt pool.

In the perifreezing zone, we find that the magnitude of the temperature forcing is $\mathcal{O}(5-30K)$ much less than Earth's seasonal forcing, on timescales of ten to a hundred days. We expect lower ice growth and melting rates than on Earth, and thinner ice equilibrium thicknesses. In general, our steady state, smoothly varying ice thickness approximation will break down in the perifreezing zone. By analogy to Earth, we expect that sea ice drift will lead to areas of mixed ice and water and a range of ice types with different thicknesses.

Sea ice drift is driven primarily by wind stress, but also by ice-ocean stresses, Coriolis forces, and the slope of the ocean surface \citep{Steele1997}.
Sea ice normally drifts at an angle of 10-40\degree \, clockwise (anticlockwise) from the wind direction due to Ekman transport in the Northern (Southern) hemispheres.
Ice drift speeds on Earth are about 1-2\% of wind speeds \citep{Thomas2010}. Given the average surface wind speed of 10s of ms$^{-1}$, this will lead to ice drift velocities of 0.1-1ms$^{-1}$, consistent with the results of \citet{Yang2020}, or 10-100 km/day. On a small spatial scale, ice movement will lead to stretches of open water, collisions, and ridging as thin forming ice sheets collide with each other. This will lead to a range of ice types forming, starting with needle-like frazil, then thin ice sheets (nilas) or thicker pancake ice depending on the winds, which in consistently freezing temperatures will coalesce into larger ice blocks.

The impact that sea ice drift will have on the large-scale sea ice distribution and general climate state depends largely on the surface wind distribution. As seen in our simulations, slow-rotating tidally-locked exoplanets will generally have surface winds which converge towards the substellar point. Sea ice will then generally drift from the ice edge region towards the substellar point, creating leads and cracks in the ice edge region, and chunks of ice slowly melting in the open ocean. The ice convergence can be expected to lead to ridging and increased ice thickness in these areas, with some corresponding divergence on the nightside where surface winds are divergent. 

In the limit where the surface winds are perpendicular to the ice shelf edge, the ice velocities we estimated above imply it would take $\mathcal{O}(10-100$ days) for a given block of ice to cross from one side of the surface perifreezing zone to another. However, just because the surface winds are generally towards the substellar point does not mean that they are always perpendicular to the ice edge. If the dominant winds instead go from the open ocean to the ice sheets -- as can be the case if these winds are mostly zonal as in Case 3 -- then we expect to see increased ice thicknesses on the ice edge, balanced by thermodynamic melting. Another case can be seen in Figure \ref{fig:surfwinds}, where there are regions of the surface where the ice floes drift mostly perpendicularly to the ice shelf edge. In this case, an ice block's residence time in the perifreezing region will be increased to several months or years, depending on the drift velocity. This is a significant extension of the time that ice floes will spend in a regime with the temperature fluctuating between above and below freezing, with small scale melting and freezing happening in the meantime. If there are suitable atmospheric conditions e.g. wind gyres at the position of the ice sheet edge, this time might be extended indefinitely, as long as the average temperatures stay cool enough to not completely melt the ice block.

Finally, we note that this largely static perifreezing zone is a result of simulating only tidally-locked planets with zero obliquity. Given that tidally-locked planets make up the majority of currently known habitable exoplanets as well as those detectable in the immediate future, we feel that this is a reasonable choice. However, while tidal dissipation will normally push these planets down to zero obliquity, this is not always the case. Planets in compact resonant systems where planet-planet interactions can remain in stable states with non-zero obliquity \citep[e.g.][]{Chen2023,Millholland2024}. Additionally, it is possible for planets to instead end up in e.g 2:1 or 3:2 resonances with their host star. Finally, in the case of fast rotating, non-tidally locked planets, we would expect a degree of non-zero obliquity. In all of these cases we would see spatially-evolving radiative forcing on a timescale similar to the orbital period. This would likely lead - for long enough orbital periods and short enough planetary radiative timescales - to large portions of the planet's ice sheet completely melting and refreezing. However, we could also expect some regions (e.g. towards the poles if in a spin-orbit resonance with low obliquity), we would expect more moderate temperature changes, leading to recurrent partial melting and freezing similar to that found in our perifreezing zone.

To summarise, if the climate conditions lead to partial dayside sea ice coverage, we expect ice formation and destruction largely centred in a `perifreezing' zone where surface temperatures fluctuate around freezing, but with sea ice drift and air temperature variability having the potential to substantially extend this zone. Depending on the surface winds around these perifreezing zones, sea ice can either drift steadily into warmer conditions, into colder conditions, or stay a long time in the perifreezing zone. Long residence timescales of hundreds of days are plausibly common, with potentially much longer residence timescales with the correct wind/ice configurations. In this perifreezing zone, the behaviour of Earth sea ice leads us to expect a complex mix of ice forms ranging from open water near the freezing point, to thin ice sheets (nilas), to thicker ice chunks formed by wind-driven collisions between smaller ice blocks.

\subsection{The potential for ice sheet margins to concentrate nutrients}
\label{sec:seaice_icemargins}

Here we outline how small-scale ice processes might lead to high concentrations of prebiotic molecules in ice shelves, and particularly on the ice sheet edges. The central mechanism which can drive nutrient concentration is the salinity enhancement of brine pockets: this can drive arbitrary concentration enhancements limited only by the final brine temperature. In the dilute limit, the freezing point of water drops by $1.86$~K per mole of dissolved solute.  The colder the water-ice mixture, the greater the fraction of H$_2$O in solid phase and the more concentrated the brine - e.g. if the coldest temperatures reached are 263~K, then an Earth-like brine's concentration will triple to about 2 mol \citep{Assur1960}. In this way, molar concentrations can be reached from possibly very dilute starting material, as has been thought to be helpful for potential Earth sea ice prebiotic chemistry \citep[e.g.][]{Miyakawa2002a, Trinks2005, Attwater2010}.

There will always be a certain amount of bioessential elements and molecules in the global ocean (even if in trace amounts), and brine formation would consistently lead to their concentration. This material will have been accreted alongside water during planet formation, or delivered by impacts or condensing from the primordial envelope following \citet{Madhusudhan2023a}. For a maximally efficient concentration of these contents of the ocean, a few conditions must be met. Solutes must not be incorporated into the ice matrix, they must stay dissolved (this depends heavily on the solutes - many prebiotic molecules are quite soluble in water, but a number of useful reactants and in particular minerals are not, as discussed in Section \ref{sec:summary}), and the brine pocket must exchange only heat with its surroundings (and not drain out of the ice). If conditions not being met, then nutrient concentration in brine will be less effective and happen less often. This is notably a risk if the overall ocean is not saline enough, as freshwater ice has few brine inclusions and particular dynamical conditions during ice formation might be required for consistent nutrient concentrations to happen. 

We can also go a step further than considering an isolated brine pocket and a unidirectional trend in concentration. One of the advantages of a stable ice margin zone is that, as shown in Figures \ref{fig:inn2nd_gridcell}, \ref{fig:out2nd_gridcell}, and \ref{fig:innpri_gridcell}, its fluctuating temperatures could allow for freeze-thaw cycling and for arbitrarily high nutrient enhancements. The following example, while fine-tuned, serves as an optimistic vision to illustrate the possibilities of the `pool-on-an-ice-block' idea on the edge of the ice sheets.

First, take an ice volume that has been affected by dynamical motions, such that it has a rough surface -- this could be either pancake ice or a thicker, larger ice keel, for example formed by surface winds driving convergent ice movement. Next, place this volume in the perifreezing zone, a mix of ice and open water. Then, hope that the features of the perifreezing zone (size, dominant wind variations, climate variability) cause the ice volume to remain for an extended period of time in this perifreezing zone with temperature at or somewhat below freezing, as was the case in Figure \ref{fig:innpri_gridcell}.
In these conditions, collisions, wind-driven spray or waves would cause seawater to splash onto the rough top of the ice volume, causing a small pool. In most cases this will either freeze out completely with the dissolved salts etc forming a crust on the top, or run off the top, or evaporate. However, in the right conditions, the pool will slowly start to freeze, and so the concentration of the pool will increase. The pool will reach a limiting concentration driven by the temperature reached.

The difference between this scenario and the brine inclusions in the ice is that additional seawater can end up in the pool,  creating wet-dry cycling. The mix of concentrated pool water and dilute seawater will dilute the pool's concentration, but increases its total reservoir of nutrients. If the ice volume ends up in colder conditions again, the pool can freeze and further concentrate. If this process is repeated several times, we would end up with a series of wet-dry cycles which in the longer term steadily increase concentrations in the pool, by allowing for a theoretically unbound supply of nutrients into a constricted volume. Conversely, local ice temperatures above freezing, if combined (as is likely) with a suitable ice floe geometry directing meltwater towards the pool, would lead to freshwater added to the pool and reducing its salinity. Significant time with local temperatures above freezing and an ice surface at the melting point would thus be unfavourable to nutrient enhancement, and instead the local ocean would ideally be a ice-water mix around the freezing point of seawater. Luckily, this is quite commonly the case in the perifreezing zone, as in Figure \ref{fig:innpri_gridcell}. 

This hypothetical process will end up limited in one of two ways. Firstly, high enough concentrations in the pool depress the mixture's freezing point such that further freezing is impossible. Whether this results in conditions suitable for prebiotic chemistry or not will depend on the ratio of initial nutrients in the global ocean.
The second limitation is the ability of the wider climate conditions to allow this repeated seawater resupply and then freezing. For this cycling to happen, the ice volume must spend an extended amount of time in regions with temperatures at or below freezing, with at least some amount of open water. 

The porosity of the ice floe is a crucial characteristic and ideally, the volume would never be in a region which goes above 0$\degree$C. For the pool to not drain into the rest of the ice, the porosity of the ice must be low, so it must have been well below freezing at some point. Some properties of melt pools seem fairly steady across different Earth conditions \citep[e.g. pond scales and interconnectednesses,][]{Popovi2018}, so we might expect that their key characteristics will carry across to exoplanets. One difference is the ice albedo, which is lower for an M dwarfs' stellar flux - pool growth is quite sensitive to the sea ice albedo \citep{Popovi2017}, so we might see faster melt pond growth when the stellar flux is high enough. In general, top melting of the ice volume might just lead to another wet-dry cycle, for example through meltwater runoff into the pool. However, significant melting is more likely to lead to meltwater flushing of the ice volume as the ice becomes permeable, expelling the denser brine downwards into the ocean. 

We view this as the most significant potential flaw with this concentration mechanism. Further simulations or observational/experimental work is required to investigate how the porosity of old, cold, sea ice (multiyear ice, in Earth terms) evolves when subject to fluctuating temperatures approaching the melting point of ice. Significant time spent with the ice top melting is very likely to lead to meltwater flushing and the ice becoming permeable. Even if the ice stays below the melting point, it is possible that brine pocket drainage starts to happen and that the prebiotic reactants are flushed into the global ocean. For a stable splash pool to survive, it would likely be necessary that the bulk salinity of the ice is lower than usual on Earth: either through a less saline ocean, or old ice which has lost most, but not all, of its initial brine inclusions.

Overall, the most favourable scenario in which all of these conditions are met involves an ice volume which starts in the thick ice sheet region, well away from the ice shelf edge. Surface winds will drive it towards the perifreezing region, where convergent ice motion leads to deformation and ridging. Once in the perifreezing region, the dominant winds are either weak, so the ice block stays mostly in place, or bring the ice volume along the edge of the perifreezing region, where there is a long stretch of wet-dry cycles during which the total seawater concentration can rise to molar levels and prebiotic chemistry can happen. We can also imagine a case in particularly favourable wind conditions or ice margin geometries, where the ice block ends up cycling between the permanently freezing ice sheet and the perifreezing zone - however, the sea ice drift patterns on most planets are unlikely to allow this.

\subsection{Numerical simulation of a drifting ice block}
Our GCM simulations include detailed, time-varying outputs of the surface climate conditions. This is very useful, as it allows us to actually numerically simulate the evolution of drifting ice floes in these conditions, and to begin to address whether the persistent wet-dry cycling invoked above may be possible. We find that the drifting ice blocks can survive significant amounts of time, and that freeze-thaw cycling does indeed happen on timescales of weeks to months.

We have developed a simplified version of the CICE sea ice code used in the ExoCAM GCM, and use this to evaluate the survival and temperature evolution of an ice block subject to a given climate forcing. Technical details of the code can be found in \citet{Hunke2008}. Simply, we keep the standard CICE thermodynamics, model only a single ice thickness and not an ice thickness distribution, and neglect any ice thickness changes due to ice drift and dynamics (as we consider a single ice floe drifting in open water). We neglect snow for simplicity - given the high albedo of snow and the conversion of snow to ice, this is a conservative approximation that hinders ice survival. We also do not explicitly simulate the briny surface pond, assuming it is small compared to the ice block.

\begin{figure*}
    \centering
    \includegraphics[width=0.95\textwidth]{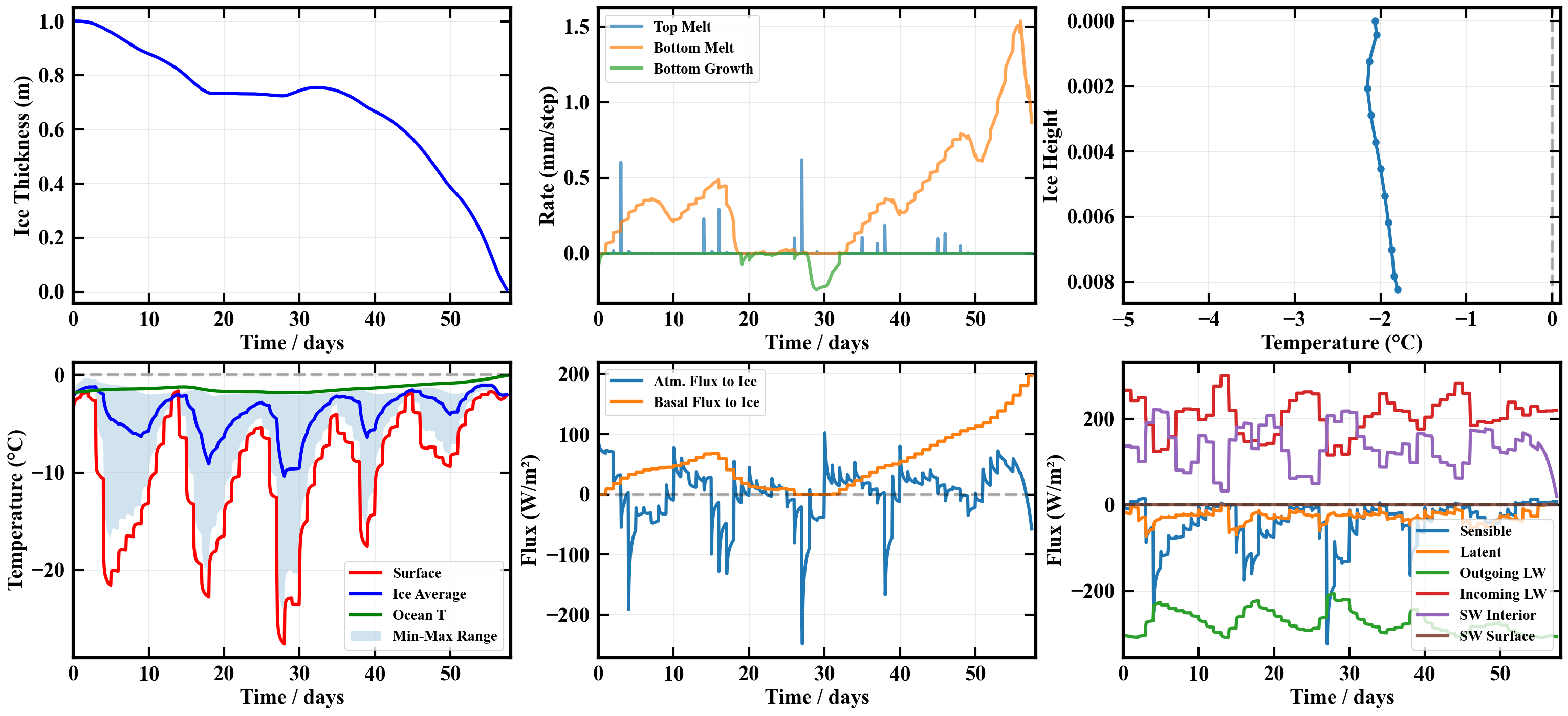}
    \caption[Case 1 ice block evolution]{The evolution of an ice block following the prevailing winds and moving directly across the perifreezing zone in Case 1. Top left: ice thickness evolution. Top centre: The ice melting and growth rates. Top right: the final pre-melting in-ice temperature profile. Bottom left: The temperature evolution of the ice block. Bottom centre: the net atmospheric and ocean fluxes to the ice. Bottom Right: decomposition of the atmosphere-ice flux into its constituents. The ice block is initialised as a 1 metre thick, snow free, ice floe which starts moving from 50 \degree $ S$, 37\degree $ W$. It moves northwards at a speed of 0.15 m/s or $\approx $ 13 km/day, until it melts completely 50 days later at around 43 \degree $ S$, 37\degree $ W$. The key driver of the ice melt is the ocean temperature, and most melt happens at the ice-ocean interface. Major atmospheric temperature variations on a timescale of about 10 days are matched by the surface reaching temperatures as cold as $-20$\degree$ C$. Note that the discretisation of some values, particularly apparent in the longwave and shortwave fluxes, is because our forcing data was recorded daily in the GCM, whereas the ice model timestep is significantly shorter than that.}
    \label{fig:inn2nd_iceblock}
\end{figure*}

\begin{figure*}
    \centering
    \includegraphics[width=0.95\textwidth]{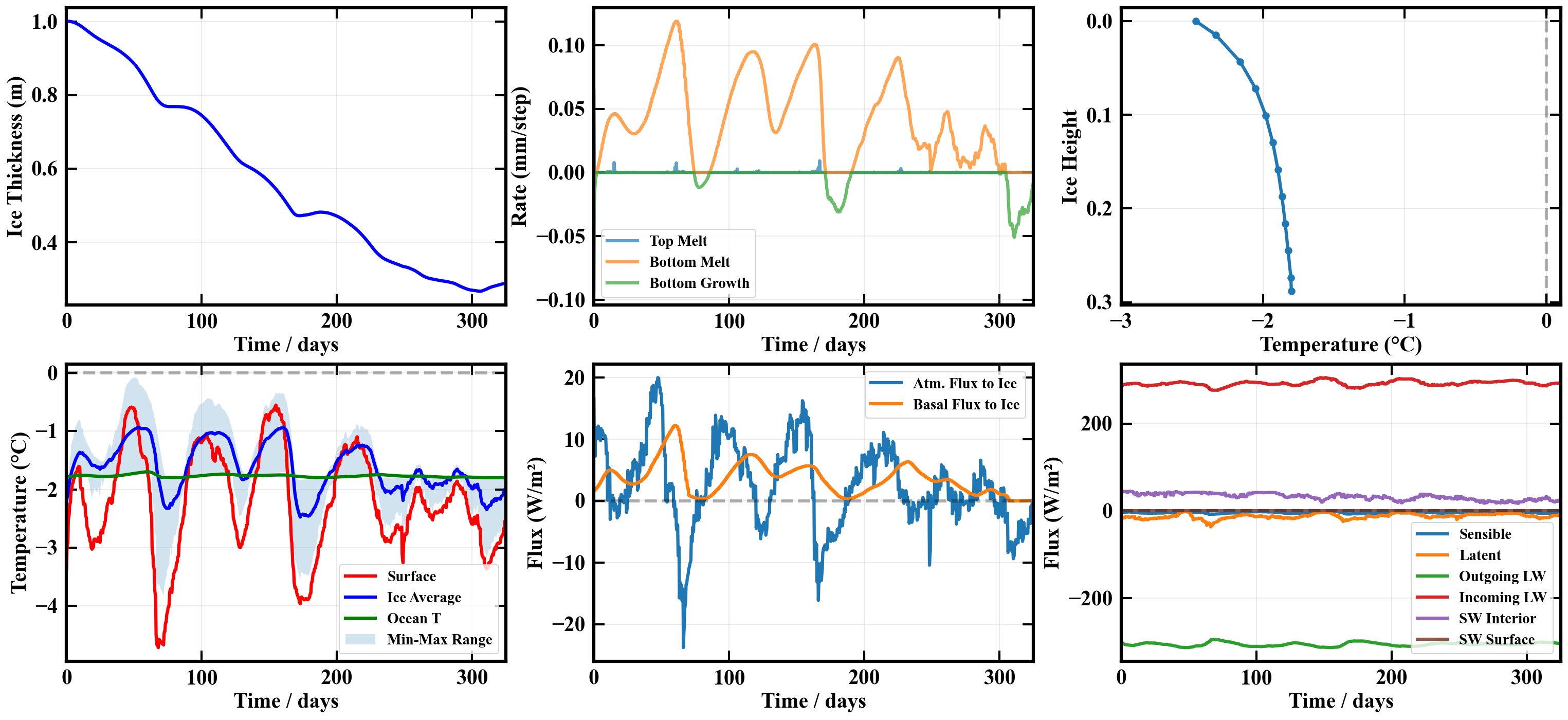}
    \caption[Case 3 ice block evolution]{The evolution of an ice block following the prevailing winds and moving directly along the edge of the perifreezing zone in Case 3. Top left: ice thickness evolution. Top centre: The ice melting and growth rates. Top right: the final pre-melting in-ice temperature profile. Bottom left: The temperature evolution of the ice block. Bottom centre: the net atmospheric and ocean fluxes to the ice. Bottom right: decomposition of the atmosphere-ice flux into its constituents. The ice block is initialised as a 1 metre thick, snow free, ice floe which starts moving from 38 \degree $ S$, 43\degree $ E$. It moves east-north-east at a speed of 0.1 m/s or $\approx $ 8 km/day, and generally slowly melts, reaching a thickness of 0.3 m 330 days later at around 34 \degree $ S$, 53\degree $ E$. Most melt is again bottom melt, and the surface varies between $-1$ and $-4$\degree$ C$.}
    \label{fig:innpri_iceblock}
\end{figure*}

To show the span of possible behaviours, we present two different cases for the ice block: one following an ice block moving directly across the perifreezing zone into the open ocean, and one following a block moving along the perifreezing edge. The first case is designed to see what the conditions are in less favourable conditions for ice block survival. We follow an ice floe starting from the centre of the Case 1 grid cell centred at 50\degree $S$, 37\degree $W$, directly to the south of the one shown in Figure \ref{fig:inn2nd_gridcell}. The ice block moves northwards, directly across the perifreezing zone, at 0.15 m/s (vs the $\approx10$ m/s average wind speed), and the climate forcings it experiences are constructed from an interpolation between the neighbouring cells. The evolution of the ice is shown in Figure \ref{fig:inn2nd_iceblock}. The ice melts over the course of about two months, during which it crosses the majority of the perifreezing zone. This gives us an indication for the melting timescale of an ice floe directly crossing the perifreezing zone. The surface stays freezing, and the ocean temperature emerges as the dominant control on ice melting: the vast majority of the melt comes from the bottom of the floe. In particular, when the ocean temperature exceeds $\approx -1$\degree $C$ and particularly as it approaches $0$\degree $C$, the basal heat flux becomes significant and the ice melts quickly in $\mathcal{O}$(10 days).

In the opposite scenario, where the prevailing winds run along the edge of the perifreezing zone, we confirm our prediction that drifting ice survives much longer, possibly even indefinitely. To test this, we follow an ice floe starting from the Case 3 grid cell at 3S\degree$ S$, 43\degree$ W$, and moving East-North-East at 0.1 m/s (vs an average wind speed of about 6 m/s) until it reaches the centre of the grid cell at 34\degree$ S$, 53\degree $E$, 330 days later. The climate forcings are again calculated from interpolating between the neighbouring grid cells, and the evolution of the ice is shown in Figure \ref{fig:innpri_iceblock}. In comparison to the Case 1 scenario, here the drifting ice survives for a year, stabilising at a thickness of about 0.3m, with bottom melting dominated. This serves as a demonstration that ice floe survival over many months could be possible in the right circumstances, which would leave a significant amount of time for nutrient concentration and prebiotic chemistry. It is worth noting that the prevailing winds at that point start to incline towards the permanent ice sheet. If ice drift follows the average winds, the ice block would start then to slowly rethicken - however, the (unsimulated) ocean circulation also affects ice drift, and this might not be the case.

The variations in ice surface temperature allow us to assess the prospects of freeze-thaw cycles and the properties of the small pool suggested in the previous section. In the Case 1 scenario, surface temperatures vary on a $\approx 10$ day timescale, commonly varying between -2\degree$ C$ and about -20\degree$ C$. This implies, roughly speaking, a saline pool concentration varying between about 1 mol and 10 mol. This pool - depending on the makeup of the ocean's dissolved solutes and how many of them are bioavailable nutrients - could be expected have a high concentration of feedstock molecules alongside strong freeze-thaw cycling, favouring prebiotic chemistry. In the Case 3 scenario, the surface temperatures vary between $-1$ and $\sim-4$\degree$ C$, implying much more moderate concentration changes, and a maximum brine concentration about about 2 mol. This would happen over longer timescales of $30-50$ days, and the ice block's freeze-thaw cycling would thus be less intense but leave more time for prebiotic chemistry to occur. We note that the length of the time in the perifreezing zone and the magnitude of the wet-dry cycling are two largely disconnected properties, and are not necessarily in tension with each other. The first results from the alignment of the ice drift directions with the perifreezing zone, and the second from the degree and timescale of atmospheric variability. Depending on the climate and atmospheric dynamics of the planet, these quantities could both be significantly different than we have found in our simulations presented here.

\subsection{Impactors}

\label{sec:impactors}

So far we have concentrated on the edge of the ice shelf, where ice formation and evolution may provide an opportunity to concentrate prebiotic nutrients from dilute seawater. In this section we investigate an second possibility: what if small impacts can deliver suitable conditions for prebiotic chemistry? The role of impactors in facilitating prebiotic chemistry has been studied previously due to its potential for creating suitable abiogenesis environments as well as delivering prebiotically important molecules \citep{Chyba1993,Osinski2020}, or even more complex configurations such as multiple impactors \citep{Anslow2025a}. However, these works have focused on the relevance of impacts on terrestrial planets, and not into planets with deep water oceans.

We focus on estimating the surface flux of impactor fragments small enough to not break through the surface ice sheet, which would then result in a locally very high concentration of nutrients. Our calculations are presented in full in Appendix \ref{app:impact_flux_calc}. Here, we summarise the key steps.

\begin{itemize}
    \item Using results from \citet{Bland2006}, we find that stony asteroids of up to $10^7$-$10^8$ kg fragment before reaching the surface, typically into a large number of small objects (a few kg at most). 
    \item Using the ice perforation model of \citet{Ross1967}, we find that an ice thickness of 40 (2.3) cm is sufficient to stop a 1kg (1g) impactor: given the ice thicknesses we find in our GCM simulation, the vast majority of small fragments will remain on the surface.
    \item Linarly extrapolating the modern-day Earth impact flux, we find that 10 kg m$^{-2}$ of small impactor fragments have been delivered to the Earth over its history.
    \item Planetary parameters such as the age, radius, gravity, surface pressure, and atmospheric composition, will all affect the impactor fragment flux, but are unlikely to significantly reduce the flux compared to Earth-like values. A non Earth-like planetary system architecture could, however, reduce the impact flux in some circumstances.
\end{itemize}

This delivery of 10 kg per m$^2$ of the planet's surface is a significant amount of material that will fragment and not perforate the ice sheet, largely contained in material in sizes between mm and decimetres and masses between grams and kilograms. Of this, a substantial fraction (2-10\%) would fall on ice in the perifreezing zone, and so be immediately subject to freeze-thaw cycling, and additional material would fall on stable ice that may drift into the perifreezing zone. These fragments may sit on top of the ice sheet for a considerable amount of time: until they are weathered away, or if the ice melts enough that the impactor falls through to the ocean. However, sea ice is a dynamic environment and we do not expect that they will survive over geological timescales, and so each new impactor will probably fall on new sea ice. We note that we do not anticipate a sharp threshold (e.g. 1 kg m$^{-2}$) above which impactor delivery becomes sufficient for prebiotic chemistry. Instead, there would be a vast number of small regions where impactors would stochastically deliver material in high concentrations. Increasing the total planetary flux of impactor fragments equates to increasing the number of different dice rolls which might lead to abiogenesis.
Whether these impactor fragments could provide promising potential locations for prebiotic chemistry depends on the asteroid composition, the impact conditions, and the atmospheric conditions. We discuss this in Section \ref{sec:prebiotic-chemistry}.

\section{Summary and Discussion}
\label{sec:summary}

We investigate the possibility that dayside sea ice could be a suitable location for prebiotic chemistry and abiogenesis. This would significantly improve the habitability of planets with deep global oceans, which otherwise pose difficulties for concentrating feedstock molecules to the required levels. This may be a relevant scenario for a large number of planets, from terrestrial planets with global oceans to volatile-rich ocean planets and Hycean planets with high-pressure ice layers between the ocean and the silicate mantle. Using the ExoCAM GCM and making reasonable assumptions about their atmospheric compositions, we show that TRAPPIST-1 e, LHS-1140 b, and K2-18 b can have dayside sea ice  -- allowing incident stellar irradiation to act as the energy source for prebiotic chemistry.

Most of the ice shelf is permanently frozen, but we find large-scale ice formation and melting at the edge of the open ocean. We find average ice thicknesses at the ocean edge of about 0.5 m, and winds that usually converge towards the substellar point, but in some cases run parallel to the ice edge or even into the ice sheet. We find that across all our simulations there is significant climate variability, with temperature variations of 5-30 K with timescales ranging from tens to hundreds of days. We confirm that the cloud distributions allow for substantial shortwave radiation on the sea ice, as long as the atmosphere's gas optical depth is not too high.

We investigate two specific mechanisms that may provide chemical feedstock in prebiotically relevant quantities, and find that in both cases we would expect them to provide significant nutrient concentrations. The first is in the increased concentration of nutrients in seawater into a `brine' during ice formation, which provides a straightforward mechanism to concentrate a dilute global ocean to the molar levels. We show that in some regions of our simulations, a meandering ice floe can undergo repeated freeze-thaw cycling, possibly further concentrating nutrients and encouraging certain prebiotic reactions. The second mechanism we consider is for impactor fragments to land on the ice surface without breaking it, providing locally very high concentrations of elemental feedstock. We show that an Earth-like impactor flux extrapolated over 4 Gyr leads to a significant total impactor fragment flux of $\sim 10$ kg per m$^2$ of the planets' surfaces which could provide recurring locations for abiogenesis.

However, as we discuss more below, it is unclear what specific pathways for prebiotic chemistry would be possible with these concentrated feedstock molecules. Significant unknowns remain in the understanding of prebiotic chemistry on Earth which makes it even harder to investigate possible pathways for novel environments as considered here. Water-ice eutectics have favourable properties for a number of prebiotic chemistry reactions, but popular prebiotic reaction networks in the Earth context do not seem applicable in this case. We do speculate that water-formamide or water-urea mixtures might produce a favourable starting point for prebiotic chemistry.

\subsection{Prebiotic Chemistry in Ice}
\label{sec:prebiotic-chemistry}
The idea of ice as a setting for prebiotic chemistry has previously been explored for both Earth and the solar system icy moons \citep{Camprubi2019}. Most of this research has been motivated by some favourable physico-chemical properties of ice. In particular, a water-ice eutectic can be very useful for dehydration and condensation reactions \citep{Hulshof1976}, which can otherwise be hard to induce in prebiotic chemistry networks. As previously discussed in Section \ref{sec:seaice_icemargins}, brine formation promotes high solute concentrations, even from very dilute starting points. The low temperatures reduce reaction speeds, but also protect molecules from hydrolytic degradation \citep[e.g.][]{Miyakawa2002a,Attwater2010}. The ice microstructure, with many small brine channels and pockets, encourages compartmentalisation as a protocellular medium \citep[e.g.][]{Monnard2008a}, which could be very useful at the later stages of prebiotic chemistry. However, the high ionic strength of dissolved salts can inhibit some reactions \citep{Trinks2005}. This means that less saline seawater may in fact be more favourable for prebiotic chemistry than Earth-like or higher salinities, although various salts can also be helpful catalysts for a number of reactions.

A series of research has examined different stages of prebiotic chemistry in the lab and shown that these are possible in water-ice mixtures. \citet{Levy2000,Miyakawa2002b,Cleaves2006} examined the evolution of frozen NH$_4$CN solutions, in one case storing a mixture at -78 \degree$C$ for 27 years (!), although we use this as an example and not to suggest that the same reactions could be at play in our case. They reported the formation of simple precursor molecules: amino acids, purines, and pyriminides, which are necessary for forming RNA and DNA bases. A number of other studies have studied polymerisation reactions in eutectic mixtures, and found that the wet-dry cycling possible considerably enhances the reaction yield. \citet{Monnard2003} found that when a dilute solution of random RNA monomers was frozen, they combined efficiently to form oligomers, and \citet{Monnard2008b} found that this was encouraged by the presence of metal ions.
\citet{Trinks2005} varied the temperature of a solution between -7\degree$C$ and -24\degree$C$ and found efficient polymerisation of adenylic acid, creating very long chains of up to 400 nucleotides.

\citet{Attwater2010} examined another type of prebiotic chemistry, and showed that ice can be a protocellar medium for RNA replication, enabling the formation of long replications products at -7\degree$C$. \citet{Attwater2013} also showed that a ribozyme undergoing in-ice evolution was able to adapt to the cold conditions and synthesise RNA sequences longer than itself -- an important step towards self-replication. In the RNA world hypothesis, the isolated RNA molecule is unstable in hot conditions, which has been an argument for a cold origin of life. The pH gradients possible in freeze-thaw cycles might also be important, as \citet{Mariani2018} showed was helpful for RNA strand separation.

\vspace{0.2cm}

However, many current prebiotic chemical theories have steps which seem inconsistent with sea ice abiogenesis. Some prominent hypotheses, such as the cyano-sulfidic scenario \citep{Patel2015} and the borate-stabilised network \citep{Benner2019}, have several steps which could work in icy conditions, such as wet-dry cycling leading to ferrocyanide salt formation. However, other steps, such as the formation of sodium or potassium cyanide require hot and dry conditions (from geothermal heating or a large impact, fiducially), so these scenarios would not be feasible on an ice sheet. 

Another prebiotic chemistry scenario is sometimes called the `warm comet pond' scenario \citep{Pearce2017}, and is relevant to our idea of impactor fragments. Here, impactor material would end in up a pond which then undergoes wet-dry cycling and prebiotic chemistry \citep[similar ideas can be found in e.g. ][]{Clark1988,Walton2024}. This is similar to our invocation of impactor fragments forming small pools on the ice surface, with the difference being that our pond sits on a bed of solid ice: this constrains the temperature and rules out additional pond-rock interactions. The possible prebiotic chemistry pathways are still similar, again with the exception of those that require high temperatures. 

However, it is worth noting that some works disfavour direct delivery of reactive prebiotic material by impactors. The presence of complex organics such amino acids and nucleobases on asteroids and comets is well established, both in space \citep{LeRoy2015, Naraoka2023,Glavin2025} and in some meteorites \citep{Kvenvolden1971,Elsila2016}. However, these organics material seems to be mostly present as kerogen-like insoluble and unreactive material \citep{Pizzarello2010,Schmitt-Kopplin2010}: it is not clear how such material might be made bioavailable. This is not definitive: there will still be delivery of some soluble organics \citep[e.g.][]{Naraoka1999,Pizzarello2012}, although possibly in trace quantities. There are also indications that very carbonaceous micro-meteorites can provide some bioavailable nutrients \citep{Dartois2013,Rojas2021}, and some impactors seem to have complex enough chemistry to support possible metabolisms \citep{Fisher2026}. However, these micro-meteorites only make up a small fraction of the impactor fragment flux we calculated in Section \ref{sec:impactors}, and so may not be common enough to serve as a realistic source of prebiotic material.

Another difficulty for impactor nutrient delivery comes from the thermal processing of the impactor material as it travels through the atmosphere. The small impactors (radii of $\approx 10^1$ m) we are considering are fairly unaffected by shock heating as they impact the atmosphere \citep{Bland2006}, so the interiors may remain at moderate temperatures and allow the survival of any organic molecules. However, the outside (up to the skin depth) will be melted \citep{Mehta2018}, and some process such as erosion or a late in-flight disintegration would be required to make the -- potentially -- more bioavailable interior accessible. Generally speaking, impactor fragments are likely to provide locally high concentrations of C-, O-, N-, Si-, and Fe-rich material at the planet's surface. However, there are good reasons to believe only a small fraction of this will be bioavailable, and so -- especially when this the only feedstock available -- it may be hard to make many additional organics. 

Overall, although many individual prebiotic reactions are possible in water-ice eutectics, the most popular pathways do not seem applicable, and the feasibility of impactor debris acting as an abiogenesis location is uncertain. We stress that this is not a severe issue: the space of theoretically possible prebiotic chemistry is vast, very little of it has been explored, and the development of life on Earth itself is an unsolved problem. The only strong conclusion we can make it that much more work is necessary to assess the feasibility prebiotic chemistry possible on sea ice. This does encourage us to provide an outline of some specific prebiotic chemistry which seems plausible in our circumstances, which we do here.

Our pool-on-an-ice-block concept provides a starting point: looking beyond the generically increased concentration of the ocean's contents, some specific prebiotically promising solutions can be made. HCN, present either as leftover from impacts \citep[e.g.][]{Todd2020} or generated in reducing exoplanet atmospheres \citep{Rimmer2019}, hydrolyses in water to form formamide (HCONH$_2$). This has a freezing point of 2.5\degree$ C$, and a sublimation-ablation process, maybe after an impact onto a thick ice sheet, could create a relatively pure formamide solution \citep{Bada2016} which would be a promising solvent for future prebiotic chemistry. Alternatively, mixing of this pond's water with urea (CO(NH$_2$)$_2$), made from reactions with either HCN or NH$_3$, form a water-urea eutectic which stays liquid until -10\degree$ C$, which could also act as a solvent for prebiotic chemistry in cold conditions. In both cases, water-formamide and water-urea mixes are known to be feasible starting points for more complex chemistry leading to amino acids and other precursor molecules \citep[e.g.][]{Menor-Salvan2020}, indicating that some plausible prebiotic chemistry pathways may exist on deep ocean worlds. 

From this starting point, we would fiducially rely on favourable properties of water-ice eutectics such as freeze-thaw cycling, compartmentalisation, and pH and eH gradients to drive the formation of amino acids, complex organics, and eventually self-replicating systems. It is possible that some salts and minerals etc may be helpful by acting as catalysts, however, the insolubility of e.g. phosphates will complicate things. In general, many solutes precipitate out at too low temperatures (such as urea at -10\degree$ C$, as above), so active prebiotic chemistry would happen more easily at temperatures closer to 0\degree$ C$. 
In this picture, impacts are valuable not for the delivery of complex organic molecules, but rather as a general source of simple prebiotic molecules such as HCN, other bioessential elements such as S or P, and possible transient sources of heat (maybe allowing for temporary melt pools on the permanent freezing ice-sheet, as discussed in Appendix \ref{app:impact_flux_calc}). They also would function as a source of material to the dilute global ocean \citep[e.g.][]{Madhusudhan2023a}.

\subsection{Likelihood of deep ocean planets allowing surface sea ice}

In this work we have shown that a range of plausible atmospheres on different planets would have surface sea ice, if they have a liquid water ocean. This does not necessarily mean that dayside sea ice is likely to take place across many or most planets with substantial volatile layers. To assess this, it is necessary to consider the factors controlling atmosphere sizes for both the primary and secondary atmospheres we are studying. We note that a full investigation of this topic is well beyond the scope of it work, but we outline a few key points and summarise the current state of research.

For secondary atmospheres, the conventional habitable zone \citep[e.g.][]{Kasting1993} allows the planet's CO$_2$ partial pressure pCO$_2$ to vary in response to instellation to maintain surface liquid water, with this feedback loop being controlled by the temperature dependence of CO$_2$ drawdown via silicate weathering. For this to happen, liquid water must be in contact with a silicate crust. This case is thus restricted to an exposed rocky surface, or an ocean shallow enough that high-pressures ices do not form at the bottom. Additionally, ocean silicate weathering must have an appropriate temperature dependence, and there must be the right temperature and pH conditions such that the carbonate condensation depth is deeper than the ocean floor \citep{Hakim2023}. The temperature dependence of seafloor weathering is an area of active research \citep{Coulter2024,Fournier-Tondreau2025,Zhu2025}, and so it is not completely clear that such a atmosphere regulation mechanism exists. If it does, whether the climate stabilises at an average temperature allowing dayside sea ice would depend on the balance of volcanic outgassing and weathering. If weathering, restricted to the seafloor, is relatively weak, then such planets might tend towards a relatively warmer state, disfavouring sea ice (assuming that the outgassing of CO$_2$ is unchanged). 

This conventional carbon cycle picture would be applicable to a subset of our "deep ocean planets", those with a silicate ocean floor. Although this presence does allow for the possibility of hydrothermal vents as alternative prebiotic chemistry locations, the relevance of these on Earth is a matter of debate \citep{Seyfried2015,Moore2021,Klein2019,McCollom2016}. In this case, sea ice would provide an additional plausible location for prebiotic chemistry without being the only option.

For planets with high-pressure ice layers, silicate weathering becomes implausible as a mechanism for controlling pCO$_2$, as carbon transport between the ocean and the silicate crust would be extremely slow if it exists at all. Instead, a series of possible mechanisms have been proposed which might regulate an atmosphere's size and allow for liquid water. It is possible that such planets naturally formed with a suitably sized atmosphere to maintain liquid water \citep{Kite2018}. CO$_2$ and CH$_4$ clathrate formation, either on surface sea ice or deeper inside the ocean, might provide a regulatory mechanism \citep{Levi2017, Levi2019}, although the range in which it is relevant might be quite restricted \citep{Ramirez2018}. Other C- and N- partitioning into various high-pressure ices can also exist \citep{Marounina2020,Vazan2022,Kovacevic2022}, which might provide stability or not. Around cooler stars, CH$_4$ generally acts as an anti-greenhouse gas, adding an extra layer of complexity and meaning that the stability of atmosphere sizes night be different for different stellar types. This question of C- and N- volatile partitioning is relevant for both primary and secondary atmospheres. It is slightly less crucial for primary atmospheres as the H$_2$ also contributes significantly to the greenhouse effect - instead, an additional requirement is that atmospheric pH$_2$ is appropriate to maintain temperate surface conditions. 

Overall it is unclear which conditions might favour the presence of surface liquid water, and even less which might favour dayside sea ice. The presence of liquid water may be quite heavily disfavoured and volatile-rich planets might commonly find themselves with steam atmospheres, depending on the initial volatile inventory and volatile partitioning. However, we can use our GCM insights to make one important point: the thicker the atmosphere, the less likely it is to have sea ice. This is because horizontal temperature differences are lower for thicker atmospheres, and so the average surface temperature needs to be near freezing for there to be regions of the surface both above and below freezing. In contrast, thinner and lower-opacity atmospheres are more likely to have dayside sea ice as they will have larger surface temperature differences. 

Thick H$_2$ atmospheres in particular would require fine-tuned instellations and orbital separations to allow partial sea ice cover, as the surface temperatures gradients can be very low. Additionally, we have seen that surface temperature variability promotes large-scale ice formation and melting, which is one of the most plausible mechanisms of nutrient concentration, and these would promote habitability. The degree of surface weather (short term climate variability) also depends on the atmosphere thickness: a thick atmosphere, both in terms of pressure and optical depth, will lead to more quiescent surface conditions. Other planetary factors such as the rotation rate may also matter, as they affect the dynamical structure and can lead to global-scale atmospheric variability.

\subsection{Outlook and future direction}
This work is an initial step setting out the theory that ice shelves on deep ocean planets may be origin-of-life locations. We have shown that dayside sea ice is possible on a range of planets, and that its behaviour and characteristics likely allow a range of plausible mechanisms to concentrate material to the levels needed for prebiotic chemistry. However, there is much more work which can and should be done to investigate this idea further and assess its viability.

One limitation in our GCM model is the slab ocean model and so the lack of ocean heat transport and sea ice drift. This means that the calculated ice thicknesses in the permanently freezing regions of the planets, notably on the nightside, are unreliable, and the sea ice distribution is not completely accurate as it can be affected by ocean heat fluxes and the movement of sea ice following wind and ocean currents  \citep[e.g.][]{Yang2020}. Our current study focuses more on the general possibility of sea ice on a range of planets and the broad span of climate conditions possible around them in a general sense, instead of delimiting precisely which planetary configurations may support dayside sea ice. Therefore, while this is not a problem for our study, it would be necessary to include this to establish definitively the likelihood of sea ice across specific ocean planets.

A more accurate GCM sea ice model would be beneficial to establish the large-scale ice drift patterns and in particular how much and how far ice drifts into warmer regions towards the substellar point, which would affect the effective size of the perifreezing zone and the duration of possible wet-dry cycling. However, there are fundamental limitations of a GCM-based approach to studying ice processes. Many of the ice dynamics that we have seen are relevant -- the opening of small scale leads, ice formation and destruction in response to large-scale climate variability, the role of small scale collisions to create ice keels, and the drift of prolonged survival of a single ice floe drifting at the ice margins -- can only be studied using parametric models in a GCM. To gain insights into these processes and see if, for example, the wet-dry cycling and drift of a single ice floe proposed in Section \ref{sec:seaice_icemargins} would actually be possible, a study on ice movement with higher-resolution models (likely with boundary conditions taken from a lower resolution model) would be necessary.

Another important aspect in establishing the likelihood of dayside sea ice on ocean planets is further study of volatile partitioning and mechanisms affecting the sizes of their atmospheres. This is relevant for H$_2$-poor atmospheres; the size of the atmosphere and hence the possibility of sea ice and liquid water more generally will depend on the partitioning of C- and N- bearing species into various reservoirs: the atmosphere/envelope, dissolved into an ocean, present as clathrates, captured into the high-pressure ice layers, or stored in the silicate mantle \citep[e.g.][]{Levi2017,Marounina2020}. Understanding how all these reservoirs co-exist and interact would help constrain the relevance of sea ice as an abiogenesis location. For H$_2$-rich atmospheres, the same mechanisms are important, but the greenhouse effect and size of the atmosphere will be mostly dominated by H$_2$ (with other gases present as secondary factors), which is more sensitive to atmospheric escape \citep{Owen2019}, and thus plausible pathways to ending up with thin (bars to tens of bars) H$_2$-rich atmospheres would still need to be established.

Even if it is established that dayside sea ice is in fact reasonably common on nearby planets, the plausibility of prebiotic chemistry in sea ice will still need to be studied in more depth. While there has been consideration of ice as an abiogenesis location \citep[see e.g.][for a review]{Menor-Salvan2012}, most prebiotic chemistry research has understandably focused on more likely scenarios on Earth, such as hydrothermal vents or seasonal lakes \citep{Preiner2020,Lane2024}. As a result, it is not clear how amenable the conditions on sea ice would be: if the bioessential elements there are present in bioavailable forms; if the required nutrients will be present in the right ratios (which will necessitate simulating the chemical composiition of the global ocean); and if the various steps on the road to a fully replicating, metabolising organism can be assembled into one coherent framework. We have made some initial steps towards answering these questions, but investigating these further provides a rich avenue of research that has the potential to significantly advance our understanding of aquaplanet habitability, and particularly to improve the prospects of abiogenesis on sub-Neptunes.

\section*{Acknowledgements}

 E.B. and N.M. acknowledge support from the UK Research and Innovation (UKRI) Frontier Grant (EP/X025179/1, PI: N. Madhusudhan) towards the doctoral studies of E.B. F.E.R. acknowledges support from the UKRI Grant UKRI1195. E.B. thanks Rick Anslow and Paul Rimmer for helpful discussions. This work was performed using resources provided by the Cambridge Service for Data Driven Discovery (CSD3) operated by the University of Cambridge Research Computing Service (www.csd3.cam.ac.uk), provided by Dell EMC and Intel. 

\section*{Data Availability}

 The default ExoCAM GCM which we build off is available at https://github.com/storyofthewolf/ExoCAM. Models outputs are available on request from the paper authors.



\bibliographystyle{mnras}
\bibliography{example} 




\appendix

\section{Calculating the fluxes of impactor fragments}
\label{app:impact_flux_calc}

Throughout its history, Earth and other solar systems bodies have been subjected to a stream of cometary and asteroid impactors \citep{Grieve1994,Abramov2013}. Many of these impactors are large and will punch straight through any surface ice sheet. However, impactors below a certain size will disintegrate before reaching the surface, and cosmic dust can simply be slowed by the atmosphere. An amount of material will thus reach the surface at sufficiently low speeds that it does not perforate any ice sheets, and so result in a locally very high concentration of nutrients.

To estimate this impactor flux, we first need to determine which size impactors will break apart before they reach the surface. We focus on stony asteroids, as these make up the majority of the impactor flux - compared to these, iron meteorites are much less likely to break up in flight, and comets much more likely. We use the results from \citet{Bland2006}, who used a `separated fragment' model to simulate a range of impactor types and sizes, with the impact velocity fixed at an appropriate Earth-like value of 18 kms$^{-1}$. They find that all stony asteroids of up to $10^7$-$10^8$ kg fragment before reaching the surface, and that they typically fragment into a large number of very small objects. A large majority of fragments are kg-mass or lower, and their vertical velocities do not exceed 50 ms$^{-1}$. In fact, 80-90\% of the original impactor mass is present as less than cm-sized objects, with an impact velocity less than 20 ms$^{-1}$. The mean delivery of mass and energy to the surface is $\leq 10^3$ kg km$^{-2}$ and $\leq 10^4$ J km$^{-2}$, consistent with the impactor material being thinly scattered over a wide area. As they find that stony impactors larger than $\sim 1$ kg disrupt, this means that on Earth the overwhelming majority of material in stony impactors smaller than $10^7$-$10^8$ kg is delivered to the surface in pieces of up to a few kg falling at their terminal velocity.

Next, to consider which thickness of sea ice stops would be punctured by which size of impactor, we use the ice perforation model of \citet{Ross1967}. This is a semi-analytic model which, given the impactor mass and radius and the ice sheet thickness, calculates the critical impact velocity at which the ice sheet will be perforated. We assume that the fragments are spherical, and that they fall at their terminal velocity, $v_t = (2 mg/\rho A C_D)^{1/2}$, where $m$ is the object mass, $\rho$ its gravity, $A$ its cross sectional area, and $C_D$ the drag coefficient which we set to 0.47, matching a rough sphere in turbulent flow. 

\begin{figure}
    \centering
    \includegraphics[width=0.48\textwidth]{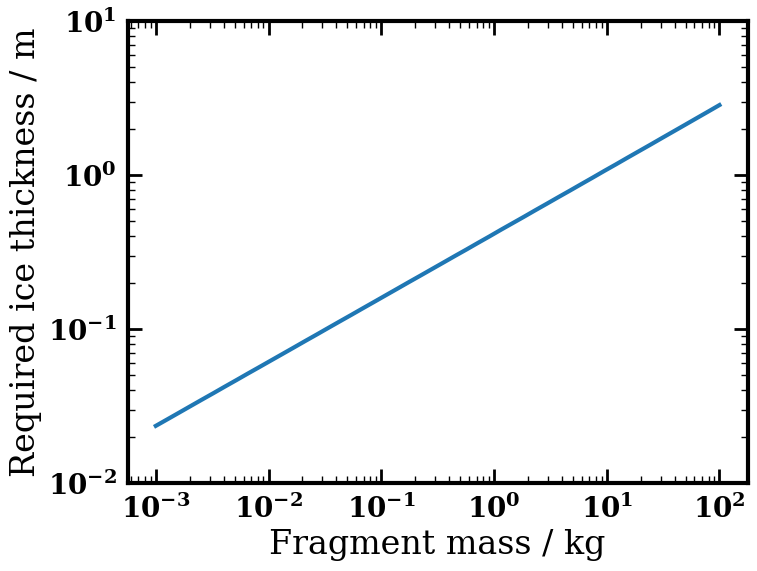}
    \caption{The dependence of the critical ice sheet thickness on the impactor mass. Bigger impactor fragments require thicker ice sheets to land on without breaking through to the ocean. The thickness-mass relation follows a power-law slope with $D\propto M^{5/12}$. Fragments of up to a kg will survive impacts onto ice with thickness $\geq$40cm.}
    \label{fig:ipds}
\end{figure}

Figure \ref{fig:ipds} shows how the maximum thickness of ice necessary to stop a given impactor fragment depends on the mass of the fragment. Given the absence of any specific length scales in our ice model, this is a self-similar solution with a power law slope of $D\propto M^{5/12}$, with $D$ the ice depth and $M$ the impactor fragment mass. It turns out that the ice thicknesses we find in our models are sufficient to stop the majority of small fragments. About 40 cm of ice is sufficient to stop a 1 kg impactor, which drops to $\sim 2.3$cm for a 1 g impactor and rises to $\sim 2.8$ m for a 100 kg impactor. The ice thickness across the whole planet is generally well over 1m, while some regions of our perifreezing zone do have ice thicknesses in the 10cm-1m range. As our impactor fragments are a few kg at most, the vast majority of small fragments -- which, as we have seen, make up the vast majority of impactor material for impactors $<10^8$ kg -- will end up stopped on the ice, and will stay there.

To calculate the flux of such impactors expected over the lifetime of our planets, we adopt Earth estimates for our fiducial values. However, we are aware that many factors such as the planet's size, atmospheric thickness, gravity, and the planetary system architecture will all affect this impactor flux. Roughly speaking, the critical mass below which an impactor will be disrupted scales linearly with the column mass of the atmosphere $(=P_s/g)$: our Case 2 would then experience a higher small impactor flux than the other two cases \citep{Anslow2025b}. A lighter atmosphere - lower $\mu$ - has little direct impact on the disruption threshold, but any impactors will have a longer passage through the atmosphere, and so will experience increased thermal processing. An increased surface gravity $g$ will also have an impact: ablation is more effective for higher impact velocities \citep[e.g.][]{Mehta2018}, likely reducing the size of the impact fragments. Conversely, it will lead to higher terminal velocities, and so increase the ice thickness needed to support a given fragment. Overall, varying the atmospheric properties away from Earth's is unlikely to significantly reduce the fragment flux, as long as the surface pressure does not significantly decrease.

The gravitationally-focused cross section of the planet scales \citep[e.g.][]{Gargaud2011} as $\sim R_p^2 (1+(v_{esc}/v_{rel})^2)$. The relative velocity between the planet and the impactor $v_{rel}$ is approximately the planet's Keplerian velocity, and the planet's mass impacts the cross section through the escape velocity $v_{esc}$. We can thus expect larger, more massive planets to have higher total impactor fluxes, although the impactor flux will be spread over a surface area also scaling as $R_p^2$, so the impactor flux in $kg/m^{-2}$ may not change very much. The overall planetary system architecture also makes a significant difference: if there are no planets exterior to the habitable zone planets we consider, and any debris belt lies far out from the star, any impactor flux will be significantly reduced \citep{Anslow2023} - however, observational constraints on the existence of these are very limited and we are not able to assess whether this is a relevant consideration. Alternatively, multi-planet systems such as Trappist-1 are expected to be efficient at scattering impactors onto inner planets \citep{Kral2018,Marino2018}, and can be expected to have greater impactor fluxes than systems without tightly packed planet chains. Our Case 1 planet, in these circumstances, might receive an increased impactor flux.

The cumulative impactor flux on Earth from the present to a given time $t$ in the past is commonly expressed as a combination of a component linearly increasing with $t$, and an exponential increase with time: $N(t)=a(\exp{bt}-1) + ct$ \citep{Marchi2009,Robbins2014}. This exponential component represents the increased impactor flux in the early solar system, and sharply increases the cumulative impactor flux before 4 Gyr. We do not include this exponential term, and only model the total impactor flux as a linear function of time. This is done firstly because the impactor flux before 4 Gyr is poorly constrained due to a lack of accurate data \citep{Zellner2017}, and contested impact rates from e.g. the lunar Highlands. Secondly, the magnitude of the early impactor flux substantially affected the planetary climate \citep{Abramov2013}, leading to (on Earth) substantial periods with steam atmospheres, dynamical sculpting of the atmosphere and/or magma oceans \citep{Zahnle2020}. These are unfavourable conditions for sea ice to form and we make the conservative assumption that conditions were not temperate enough for the impactors to lead to prebiotic chemistry. As such, we consider modern Earth-like impactor fluxes out to 4 Gyr ago, neglecting the increased impactor flux at earlier times. This age would vary for different exoplanet system - for example, of our example case, K2-18 is roughly 3 Gyr old \citep{Sairam2025} and TRAPPIST-1 about 8 Gyr old \citep{Burgasser2017}, so this could change the final cumulative flux by a factor of a few.

For the size dependence and overall occurrence rate of the modern Earth top-of-atmosphere impactor flux, we take the mass distribution from \citet{Bland2006}: for $3<m<1.7\times 10^{10}$ kg, $\log{N} = -0.926 \log{m} + 4.739$. $N(m)$ is the number of impactors with a mass $>m$ incident on Earth per year. We note that using different impactor flux parametrisations, such as the time dependence from \citet{Marchi2009} or \citet{Robbins2014} alongside a $n(D) \propto D^{-2}$ cumulative size frequency distribution \citep{Grieve1992,Hughes2000} gives similar results. We neglect the increase in impactor flux at early times \citep{Marchi2009,Robbins2014}. The magnitude of the early impactor flux substantially affected the planetary climate \citep{Abramov2013}, leading to (on Earth) substantial periods with steam atmospheres, dynamical sculpting of the atmosphere and/or magma oceans \citep{Zahnle2020}. These are unfavourable conditions for sea ice to form and we make the conservative assumption that conditions were not temperate enough for the impactors to lead to prebiotic chemistry. As such, we consider modern Earth-like impactor fluxes out to 4 Gyr ago, neglecting the increased impactor flux at earlier times.

This impactor distribution results in a total flux of $1.5\times10^6$ kg per year of impactor material between 3 and $10^7$ kg. Extrapolated over our timespan of 4 Gyr, and assuming it is distributed equally over the Earth's surface, this corresponds to $10$ kg m$^{-2}$. This is material that will fragment and not perforate the ice sheet, largely contained in material in sizes between mm and decimetres and masses between grams and kilograms. These fragments may sit on top of the ice sheet for a considerable amount of time: until they are weathered away, or if the ice melts enough that the impactor falls through to the ocean. However, we do not expect that they will survive over geological timescales, and so each new impactor will probably fall on new sea ice.

One final question worth examining in this section is the possibility of these impactor fragments creating melt pools as they land on the ice. The maximum possible melting can be estimated from comparing the kinetic energy of the impactor, $mv^2/2$, to the energy required to melt $m_I$ of ice at temperature $T_I$, $E_{melt} = m_I(L_f + T_I \times C_{p,ice})$. For a 1 kg impactor, we get a maximum $\approx 10$ ml of melt, rising to $\approx 200$ ml for 10 kg and $\approx 5$ L for 100 kg impactors. The kinetic energy of low velocity impactors is fairly efficiently converted into thermal energy \citep{Rittel2017, Vazquez-Fernandez2019, Kositski2021}, which should favour melt formation. On the other hand, for melt to be produced the incoming kinetic energy needs to be concentrated into an ice volume smaller than the cross section of the impactor, so any melt layers produced would be thin (mm to a few cm at most). We can thus conclude that some melt pools are possible, especially for larger impactors on thicker ice, but that any melt pools would be relatively small and so short-lived.

\bsp	
\label{lastpage}
\end{document}